\documentclass[a4paper,11pt]{article}
\pdfoutput=1 
\usepackage{amsmath, amssymb}
\DeclareUnicodeCharacter{2A7E}{\gtrsim}
\usepackage{jcappub} 
\usepackage[T1]{fontenc} 
\usepackage{multirow}
\usepackage{aas_macros}
\usepackage{orcidlink}
\usepackage{subcaption}
\usepackage{todonotes}
\usepackage{academicons}
\usepackage{xcolor}
\usepackage{soul}
\usepackage{booktabs}
\sethlcolor{yellow}
\definecolor{orcidlogocol}{HTML}{A6CE39}
\usepackage{caption}

\title{\boldmath{A Model-Independent Precision Test of General Relativity using LISA Bright Standard Sirens}}
\author{Samsuzzaman Afroz \orcidlink{0009-0004-4459-2981},}
\author{Suvodip Mukherjee \orcidlink{0000-0002-3373-5236}}

\affiliation{Department of Astronomy and Astrophysics, Tata Institute of Fundamental Research, 1, Homi Bhabha Road, Mumbai, 400005, India}

\emailAdd{samsuzzaman.afroz@tifr.res.in}
\emailAdd{suvodip@tifr.res.in}

\abstract{The upcoming Laser Interferometer Space Antenna (LISA), set for launch in the mid-2030s, will enhance our capability to probe the universe through gravitational waves (GWs) emitted from binary black holes (BBHs) across a broad range of cosmological distances. LISA is projected to observe three classes of BBHs: massive BBHs (MBBHs), extreme mass-ratio inspirals (EMRIs), and stellar mass BBHs. This study focuses on MBBHs, which are anticipated to occur in gas-rich environments conducive to producing powerful electromagnetic (EM) counterparts, positioning them as excellent candidates for bright sirens. By combining GW luminosity distance measurements from these bright sirens with Baryon Acoustic Oscillation (BAO) measurements derived from galaxy clustering and sound horizon measurements from the Cosmic Microwave Background (CMB), and spectroscopic redshift measurements from observations of the electromagnetic (EM) counterpart, we propose a data-driven model-independent method to reconstruct deviations in the variation of the effective Planck mass (in conjunction with the Hubble constant) as a function of cosmic redshift. Using this multi-messenger technique, we achieve precise measurements of deviations in the effective Planck mass variation with redshift (z), with a precision ranging from approximately $2.4\%$ to $7.2\%$ from redshift $z=1$ to $z=6$  with a single event. Additionally, we achieved a measurement of the Hubble constant with a precision of about $1.3\%$, accounting for variations in the effective Planck mass over 4 years of observation time ($T_{\mathrm{obs}}$). This assumes that EM counterparts are detected for $75\%$ of the events. This precision improves with observation time as $T_{\mathrm{obs}}^{-1/2}$. This approach not only has the potential to reveal deviations from General Relativity but also to significantly expand our understanding of the universe's fundamental physical properties.}

\begin{document}
\maketitle
\flushbottom
\section{Introduction}

The General Theory of Relativity (GR) has stood as a cornerstone of our understanding of gravity for over a century. It elegantly explains how massive objects warp the fabric of spacetime, causing other objects to move along curved trajectories. However, in extreme astrophysical environments such as black holes and the early universe, the GR faces challenges in its applicability \cite{thorne1995gravitational, sathyaprakash2009physics}. Recently, a groundbreaking development occurred with the direct detection of gravitational waves (GWs) by the LIGO-Virgo collaboration \cite{abbott2016observation, abbott2016gw151226, scientific2017gw170104, abbott2017gw170817, goldstein2017ordinary}. This achievement has opened a new avenue for testing GR, particularly in the context of compact binary systems existing in relativistic regimes. These compact binaries offer unprecedented opportunities to delve into the fundamental nature of gravity across a vast range of mass scales and cosmological distances. The evolution of ground-based GW observatories such as LIGO \cite{aasi2015advanced}, Virgo \cite{acernese2014advanced}, KAGRA \cite{akutsu2021overview}, and upcoming advancements such as LIGO-Aundha (LIGO-India) \cite{saleem2021science}, Cosmic Explorer (CE) \cite{reitze2019cosmic}, and the Einstein Telescope (ET) \cite{punturo2010einstein}, has significantly bolstered our capacity to detect these waves. These observatories primarily focus on the Hertz to kiloHertz range and have observed approximately 100 signals from neutron star and black hole coalescences with expectations of more discoveries in the upcoming observation run \cite{KAGRA:2021vkt,Nitz:2021zwj,Venumadhav:2019lyq}.

However, it is the forthcoming space-based observatory, the Laser Interferometer Space Antenna (LISA)\cite{Colpi:2024xhw,Robson:2018ifk,LISACosmologyWorkingGroup:2022jok}, that is set to revolutionize our understanding by observing GWs from binary black holes (BBHs) over unprecedented cosmological distances. The mission has recently been adopted by ESA with a possible launch by 2036.

LISA is designed to explore the milliHertz frequency range, which is rich with sources such as massive BBHs (MBBHs) \cite{Klein:2015hvg}, extreme mass ratio inspirals (EMRIs) \cite{Glampedakis:2002cb,Babak:2017tow}, and stellar mass BHBs \cite{DelPozzo:2017kme} each with unique observational characteristics and expected in different redshift ranges. This ability to probe such diverse astrophysical phenomena is crucial because ground-based detectors are hindered by unavoidable noise artifact, which prevents them from accessing this lower frequency domain. Scheduled for launch in the mid-2030s, LISA will detect these sources up to very high redshifts with unparalleled precision \cite{2017arXiv170200786A}. Its ability to observe with unique observational characteristics across different redshift ranges positions it as a pivotal cosmological tool \cite{LISACosmologyWorkingGroup:2019mwx}. This extensive range allows for the mapping of the universe's expansion from local to very distant realms. Notably, MBBHs, which are anticipated to form in gas-rich environments, may generate significant electromagnetic (EM) counterparts through various processes such as jets, disk winds, or accretion \cite{Caprini:2016qxs,Palenzuela:2010nf,OShaughnessy:2011nwl,Moesta:2011bn,Okamoto:2019zgf,Meier:2000wk}. The detection of these EM emissions not only enables precise localization of MBH events in the sky but also makes them exceptionally valuable as bright sirens for gravitational studies. The combination of GWs and EM observations could lead to breakthroughs in multi-messenger astronomy, further enhancing our understanding of the universe's structure and evolution. 

Along with these, the investigation of modified gravity theories can gain insights through the propagation speed and luminosity distance of GWs and EM signals. Various factors, including the mass of gravitons, a frictional term related to the changing Planck mass, and anisotropic source terms, influence deviations from expected results \cite{deffayet2007probing, saltas2014anisotropic, nishizawa2018generalized, belgacem2018gravitational, belgacem2018modified, lombriser2016breaking, lombriser2017challenges}. Accurate measurements of these deviations provide a framework for rigorously testing alternate gravitational theories. Utilizing bright sirens is a particularly effective method for evaluating parameters like the GWs propagation speed, the mass of gravitons, and frictional effects denoted by $\mathrm{\gamma(z)}$ (see in Equation.\ref{eq:main}). The observation of an EM counterpart roughly 1.7 seconds following the binary neutron star merger GW170817 enabled the establishment of stringent limits on the speed of GWs and the graviton's mass \cite{abbott2017gw170817, abbott2017gravitational, abbott2019tests}. Some theories suggest variations in GWs speed, especially within frequency bands monitored by LISA, contrasting with those covered by LIGO \cite{deRham:2018red}. For the purpose of this study, we assume that the speed of GWs is equivalent to that of light.

The literature offers various model-dependent parameterizations of the friction term $\mathrm{\gamma(z)}$, including the popular ($\mathrm{\Xi_0, n}$) model \cite{belgacem2018modified}, which has been shown to be inferable with LVK \cite{Mukherjee:2020mha, leyde2022current} and LISA \cite{Baker:2020apq}. This particular parameterization utilizes two positive parameters, $\mathrm{\Xi_0}$ and $\mathrm{n}$, where a $\mathrm{\Xi_0}$ value of 1 aligns with GR. Various models of modified gravity propose different behaviors for this frictional term under diverse conditions. Scalar-Tensor theories, for example, introduce a scalar field that significantly influences cosmological dynamics. Notable within this category are Brans–Dicke theory \cite{brans1961mach}, f(R) gravity \cite{hu2007models}, and covariant Galileon models \cite{chow2009galileon}, all part of the broader Horndeski framework. Extending beyond Horndeski, Degenerate Higher Order Scalar-Tensor (DHOST) theories \cite{frusciante2019tracker, crisostomi2016extended, achour2016degenerate} represent a comprehensive category of scalar-tensor models that propagate a solitary scalar degree of freedom along with the massless graviton's helicity-2 mode. Other notable theories include f(Q) gravity, f(T) gravity \cite{cai2016f}, minimal bigravity \cite{de2021minimal}, and models incorporating extra dimensions \cite{dvali20004d, yamashita2014mapping, corman2021constraining}.

These theories demonstrate varying deviations from GR that are not fully captured by the power-law $\mathrm{\Xi_0}$ and $\mathrm{n}$ parameterization, particularly at the extremes of the redshift spectrum, where the fitting formula for $\mathrm{\gamma(z)}$ may lose accuracy \cite{deffayet2007probing}. Moreover, the integrated effect of $\mathrm{\gamma(z)}$ in the real universe may deviate from a simple power-law, underscoring the necessity for model-independent methods to test GR and measure $\mathrm{\gamma(z)}$ based on observational data. To address these issues, we propose a novel model-independent approach through a function designated as $\mathrm{\mathcal{F}(z)}$, defined in Equation \ref{eq:f(z)}. This function is crafted to quantify any departures from GR by comparing GWs luminosity distances $\mathrm{D_{L}^{GW}(z)}$ with their EM counterparts $\mathrm{D_L^{EM}(z)}$. In the context of standard GR, $\mathcal{F}(z)$ should theoretically equal 1, providing a clear metric to gauge discrepancies attributable to alternative theories of gravity.

\begin{figure*}
\centering
\includegraphics[height=6.0cm, width=13.0cm]{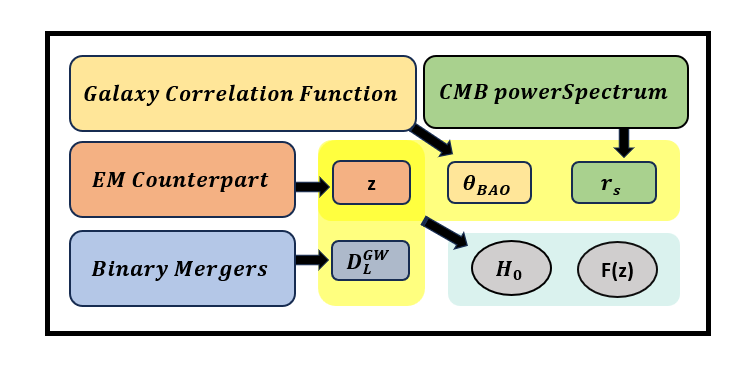}
\caption{This diagram outlines the methodology for measuring three critical length scales: the sound horizon ($\mathrm{r_s}$), the BAO scale ($\mathrm{\theta_{BAO}}$), and the GW luminosity distance ($\mathrm{D_L^{GW}}$). These scales are derived from distinct observational methods, including CMB measurements, galaxy correlation studies, and detections of GWs. The interplay between the sound horizon and the BAO scale establishes the EM angular diameter distance ($\mathrm{D^{EM}_a}$). Utilizing the distance duality relation, these measurements facilitate a redshift-dependent empirical verification of GR, as expressed in Equation \ref{eq:f}. This schematic provides a visual framework for understanding the multi-messenger approach in extracting these scales using LISA's bright sirens.}
\label{fig:motivation}
\vspace{-0.5cm}
\end{figure*}

This paper is structured as follows: In Section \ref{sec:GWprop}, we explore the propagation of GWs beyond GR. We then introduce our proposed model-independent, data-driven reconstruction of the variation in the effective Planck mass (frictional term). In Section \ref{sec:Mocksamples}, we focus on the astrophysical population of MBBHs. Section \ref{sec:EMcounterparts} addresses the identification of EM counterparts of LISA sources. Section \ref{sec:BAO} offers a detailed discussion on the model-independent measurement of the BAO scale from the galaxy power spectrum. In Section \ref{sec:Forecast}, we elaborate on the formalism underpinning our work. Subsequently, Section \ref{sec:result} presents the main findings of our methodology. Finally, Section \ref{sec:conclusion} summarizes our key discoveries and discusses future prospects.

\section{Non-parametric reconstruction of deviation from General Relativity: Formalism}
\label{sec:GWprop}

The mathematical representation of GWs propagation in spacetime, as described by GR, is given by
\begin{equation}
\mathrm{h_{(+,\times)}^{''} + 2\mathcal{H}h_{(+,\times)}^{'} + c^2k^2h_{(+,\times)} = 0},
\end{equation}
where $\mathrm{h_{(+,\times)}}$ represents the strain of the GWs in the plus ($+$) and cross ($\mathrm{\times}$) polarizations, the prime denotes a derivative with respect to conformal time ($\mathrm{\eta}$), $\mathrm{k}$ is the wavenumber of the GWs, and $\mathrm{\mathcal{H}}$ is the Hubble parameter in comoving coordinates. This equation forms the basis for analyzing GWs propagation and testing the validity of GR. Alterations to this equation in the context of modified gravity theories are expressed as 
\begin{equation}
\mathrm{h_{(+,\times)}^{''} + 2(1-\gamma(z))\mathcal{H}h_{(+,\times)}^{'} + (c_{GW}^2k^2+m_{GW}^2a^2)h_{(+,\times)} = a^2\Pi_{(+,\times)}},
\label{eq:main}
\end{equation}
where $\mathrm{\gamma(z)}$ represents a frictional term reflecting the influence of modified dynamics, $\mathrm{c_{GW}}$ denotes the speed of GWs propagation, $\mathrm{m_{GW}}$ is the graviton mass, $\mathrm{a}$ is the scale factor, and $\mathrm{\Pi_{(+,\times)}}$ is the anisotropic stress term.

Under standard GR assumptions, all the additional parameters such as $\mathrm{\gamma(z)}$, $\mathrm{m_{GW}}$, and $\mathrm{\Pi_{(+,\times)}}$ are set to zero, except for $\mathrm{c_{GW}}$, which is equal to the speed of light (denoted by $\mathrm{c}$). When testing the deviations from GR, these parameters are crucial to ascertain any potential discrepancies. Observations from the GW170817 event have placed strict limits, confirming that the graviton mass ($\mathrm{m_{GW}}$) is effectively zero, and that the speed of GWs propagation ($\mathrm{c_{GW}}$) aligns with the speed of light \cite{abbott2017gravitational, abbott2019tests}. Although some theories suggest variations in the speed of GWs outside the LVK observational frequency range \cite{deRham:2018red}, for the purpose of this analysis, we maintain the assumption that the speed of GWs is equivalent to that of light. Furthermore, changes in the anisotropic stress term ($\mathrm{\Pi_{+,\times}}$) can affect the phase of binary waveforms. Recent observations, especially of GW events like GW150914 and GW151226, have begun to constrain these potential deviations, although the constraints remain relatively broad \cite{abbott2016binary}. The friction term, denoted as $\mathrm{\gamma(z)}$, significantly affects the amplitude of GWs from cosmological sources. This alteration suggests that the luminosity distance to GWs, measured via standard sirens, could differ from distances determined through EM probes like standard candles, CMB, or BAO. This difference potentially serves as a distinct indicator of modified gravity theories. Hence, the GWs luminosity distance, $\mathrm{D_{L}^{GW}}$, is linked to the EM luminosity distance, $\mathrm{D_{L}^{EM}}$, through an exponential relationship with the exponent being the integral of $\mathrm{\gamma(z')/(1+z')}$ \cite{belgacem2018gravitational}
\begin{equation}
\mathrm{D_{L}^{GW}(z) = \exp\left(-\int dz'\frac{\gamma(z')}{1+z'}\right) D_L^{EM}(z)}.
\end{equation}
Our focus is primarily on tensor perturbations, which are captured in the ratio $\mathrm{D_{L}^{GW}(z)/D_L^{EM}(z)}$. We explore a non-parametric reconstruction of this ratio as a deviation from GR
\begin{equation}
\mathrm{D_{L}^{GW}(z)={\mathcal{F}}(z)D_L^{EM}(z)},
\label{eq:f(z)}
\end{equation}
here, $\mathcal{F}(z)$ encapsulates any deviations from GR as a function of redshift, aiding in the identification of modified gravity effects through observed GWs and EM luminosity distances introduced in the paper\cite{Afroz:2023ndy} . Alternatively, a parametric approach that employs $\mathrm{\Xi_0}$ and $\mathrm{n}$ is used to describe a power-law modification in GW propagation over redshift\cite{Belgacem:2018lbp,Leyde:2022fsc,Baker:2020apq}
\begin{equation}
\mathrm{D_{L}^{GW}(z)/D_L^{EM}(z) = \Xi_0 + \frac{1-\Xi_0}{(1+z)^n}}.
\end{equation}

This method, including both non-parametric and parametric forms, facilitates a thorough investigation of deviations from GR and dark energy's equation of state across various redshifts. Future sections will detail results for $\mathcal{F}(z)$ both independently and in conjunction with the Hubble constant, allowing for adjustments based on the sound horizon's precise estimation. GWs provide a unique method for measuring luminosity distances in binary systems, denoted as $\mathrm{D_{L}^{GW}}$. While standard $\mathrm{\Lambda}$CDM cosmology also estimates these distances, it depends on model-specific assumptions. In contrast, for a model-independent, data-driven inference of luminosity distance, we utilize calculations of EM luminosity distance derived from various length scales.

\textit{Data-driven inference of EM luminosity distance:} We employ a data-driven methodology to determine the EM luminosity distance, which leverages measurements of the angular diameter distance, $\mathrm{D^{EM}_a(z)}$, obtained from the angular peak position of the BAO ($\mathrm{\theta_{BAO}(z)}$) observed in large scale structures across various redshifts \cite{peebles1973statistical}. This approach utilizes the Etherington's distance duality relation, which establishes a fundamental connection between the EM angular diameter distance $\mathrm{D^{EM}_a(z)}$ and the EM luminosity distance $\mathrm{D^{EM}_L(z)}$. The distance duality relation is given by: $\mathrm{D^{EM}_L(z) = (1+z)^2 D^{EM}_a(z)}$. This relation is derived from the fact that in a universe described by the cosmological principle (isotropy and homogeneity), the amount of light emitted from a source spreads out over an increasing area as it travels through the expanding universe. The EM angular diameter distance $\mathrm{D^{EM}_a(z)}$ measures how large an object appears on the sky for a given physical size, while the EM luminosity distance $\mathrm{D^{EM}_L(z)}$ relates to how bright an object appears based on its intrinsic luminosity. The factor of $\mathrm{(1+z)^2}$ arises because of the cosmological redshift $\mathrm{z}$, which affects both the wavelength of the photons and the flux of the light received. Specifically, one factor of $\mathrm{(1+z)}$ accounts for the redshift stretching the wavelength, thereby reducing the energy per photon, and the second factor of $\mathrm{(1+z)}$ accounts for the time dilation effect, which reduces the rate at which photons are received. Thus, this relation allows us to convert measurements of the EMA angular diameter distance, which are typically easier to obtain from observations such as the BAO, into the EM luminosity distance needed for cosmological analyses.

The BAO peak position, as identified in the galaxy two-point correlation function, provides a direct observational basis for this calculation. The associated angular scales connect to the comoving sound horizon up to the redshift at the drag epoch ($\mathrm{z_d \sim 1020}$) through the relationship
\begin{equation}
\mathrm{r_s=\int_{z_d}^{\infty}\frac{dz\,c_s(z)}{H_0\sqrt{\Omega_m(1+z)^3+\Omega_\Lambda + \Omega_r(1+z)^4}}}.
\end{equation}
Substituting $\rm{D_L^{EM}}$ in Equation \eqref{eq:f(z)} with $\theta_{\rm BAO}(z)$ and applying the distance duality relation for EM observations, we define the function \cite{mukherjee2021testing, Afroz:2023ndy}
\begin{equation}\label{eq:f}
\mathcal{F}\mathrm{(z)=\frac{D_L^{GW}(z)\theta_{BAO}(z)}{(1+z)r_s}}.
\end{equation}
Here, the comoving sound horizon at the drag epoch, $\mathrm{r_s}$, is approximated to be about 1.02 times the comoving sound horizon at the redshift of recombination ($\mathrm{z_* = 1090}$), which is $\mathrm{r_d \approx 1.02r_*}$. This factor $\mathrm{r_*}$ is derived from CMB observations, specifically from the position of the first peak in the CMB temperature power spectrum, expressed as $\mathrm{\theta_*= r_*/(1+z_*)D^{EM}_a(z_*)}$ \cite{spergel2007three, hinshaw2013nine, aghanim2020planck}.

In this framework, $\mathrm{D_L^{GW}(z)}$ is determined from GWs events, $\mathrm{\theta_{BAO}}$ is extracted from galaxy two-point correlation functions, and $\mathrm{r_s}$ is deduced from CMB temperature fluctuations. The redshift of GWs sources can be inferred from their EM counterparts, typically observable in events involving bright standard sirens. This methodology is illustrated by a schematic diagram in Figure \ref{fig:motivation}. Consequently, the combination of these measurable quantities should conform to a consistent value across all redshifts if GR and the standard $\mathrm{\Lambda}$CDM cosmological model hold true. However, any deviations from these established theories could be identified through changes in these measurements across different redshifts. It is crucial to note that while $\mathrm{r_s}$ is model-dependent and varies above $\mathrm{z_d=1020}$, it remains constant at lower redshifts. Thus, any errors in the determination of $\mathrm{r_s}$ would affect the absolute scale of $\mathrm{\mathcal{F}(z)}$, but not its redshift-dependent trend.

Future discussions will present results considering $\mathrm{\mathcal{F}(z)}$ both independently and jointly with the Hubble constant $\mathrm{H_0}$, allowing for corrections based on potential inaccuracies in the estimated sound horizon due to erroneous Hubble constant values. This analysis supports the exploration of deviations from GR and the canonical dark energy equation of state ($\mathrm{w = -1}$) using a variety of observational probes. Previous studies \cite{mukherjee2021testing} have explored parametric deviations from GR; however, our non-parametric approach using $\mathrm{\mathcal{F}(z)}$ enables a detailed, redshift-specific reconstruction of deviations from GR through multi-messenger observations.

\section{GW mock samples for LISA}
\label{sec:Mocksamples}
\subsection{Modelling the astrophysical population of LISA source}
The formation and evolution of MBBHs and their co-evolution with host galaxies pose fundamental questions in astrophysics, with implications for understanding the regulation of star formation and the growth of galaxies \cite{Croton:2005hbr,sharma2020black}. While active galactic nuclei (AGN), powered by growing MBBHs, are believed to play a crucial role in these processes, many aspects of MBBHs growth and AGN feedback mechanisms remain poorly understood \cite{mcconnell2013revisiting,Wandel:1999xy}. One significant challenge in studying MBBHs is related to the difficulty of observing their early evolution at high redshifts. Traditional observational methods are limited to detecting only the most luminous and rapidly growing MBBHs. GWs provide a promising avenue for probing the history of MBBHs, as their propagation through the universe is essentially unobstructed \cite{Barausse:2020mdt}. The future space-based GWs detector LISA will be particularly instrumental in this area. These binaries lie within the LISA band, and LISA's capabilities will allow for observations deep into high redshifts, shedding light on the formation and evolution of MBBHs \cite{LISA:2022yao}. As a consequence, our understanding of the formation and evolution of MBBHs currently heavily relies on theoretical models. Semi-analytic models of galaxy and MBBHs formation and evolution serve as powerful tools for this\cite{Somerville:2008bx,Barausse:2012fy,Sesana:2014bea,ricarte2018observational,Parkinson:2007yh}. These models link the hierarchical formation of dark matter halos to the evolution of galaxies and their MBBHs, capturing a wide range of galaxy formation physics \cite{Volonteri:2007ax,volonteri2009journey,Barausse:2012fy,Bonetti:2018tpf}. In these models, the formation of MBBHs begins with the merger of two galaxies embedded in their dark matter halos. The subsequent evolution of MBBHs proceeds through various stages, from separations of 10s-100s kpc down to a few hundred pc, and eventually to separations of $\sim$ 0.001–0.01 pc, where GWs emission drives the MBBHs to coalesce. The dynamical evolution of MBH binaries is influenced by interactions with their stellar and gaseous environments, as well as with other MBBHs through three/four-body interactions \cite{Bonetti:2017lnj,Bonetti:2017dan}. However, fully hydrodynamic simulations, which incorporate the complex physics involved in MBBHs evolution, often suffer from poor resolution and are constrained to relatively small volumes, resulting in limited statistical power \cite{DiMatteo:2005ttp,Vogelsberger:2014kha,Schaye:2014tpa,Volonteri:2016uhr,tremmel2017romulus,Pontzen:2016wwf,Nelson:2019jkf,ricarte2019tracing}. Additionally, these simulations are computationally very demanding. While cosmological simulations have been employed to study the evolution of unresolved MBBHs, they generally necessitate significant post-processing and involve considerable simplifications. The formation and evolution of MBBHs and their co-evolution with host galaxies rely heavily on several key ingredients. These include the initial mass function of the MBH seeds, the delays between galaxy mergers and MBBHs mergers. These factors introduce complexities in modeling the dynamics and growth patterns of MBBHs. The literature presents a variety of seed models, delay models, reflecting the diverse theoretical approaches to these phenomena \cite{Somerville:2008bx,Barausse:2012fy,Sesana:2014bea,ricarte2018observational,Parkinson:2007yh,Hirschmann:2013qfl,pfister2019erratic}.

\texttt{Seed Model:} There are mainly two classes of seed models in the semi-analytical modeling of MBBHs formation: the Light-Seed (LS) and Heavy-Seed (HS) models. The LS model suggests that MBH seeds are remnants of Population III stars from high-redshift, low-metallicity environments, with masses centered around 300 solar masses and a standard deviation of 0.2 dex. These seeds are located in the most massive halos formed from the collapse of 3.5-sigma peaks in the matter density field at redshifts greater than 15, allowing for mildly super-Eddington accretion rates. In contrast, the HS model posits that MBH seeds form through the collapse of protogalactic disks due to bar instabilities, typically reaching masses of approximately $\mathrm{10^5}$ solar masses, and are found in similar high-redshift halos but with accretion rates capped at the Eddington limit. The choice between these models affects the formation and evolution of MBBHs, impacting the observed event rates and characteristics of MBBHs mergers. This differentiation is crucial for understanding the astrophysical processes leading to MBBHs and their subsequent mergers \cite{bieri2017outflows,Granato:2003ch,Barausse:2020mdt,Lapi:2013ira,ricarte2019tracing}. Recent observations by the James Webb Space Telescope (JWST)\cite{Gardner:2006ky} provide compelling evidence favoring the HS model \cite{Yue_2024}. JWST has detected extremely luminous quasars at redshifts greater than 7, suggesting the presence of very massive black holes already in the early universe\cite{2023ApJ...953L..29L}. The inferred masses of these early quasars are more consistent with the higher initial masses predicted by the HS model, as the rapid formation and growth required to reach such masses challenge the LS model’s predictions. These findings indicate that massive black hole seeds might have formed through more efficient, direct collapse mechanisms as proposed by the HS model\cite{Jeon:2024iml}.

\texttt{Delay Models:} In the study of MBBHs formation, time delays play a crucial role by affecting the evolution and eventual merger of these systems. These delays stem from various mechanisms, including dynamical friction, galaxy mergers, and the internal dynamical evolution of MBBHs within galaxies \cite{Barausse:2012fy,Sesana:2014bea,Antonini:2015sza,Klein:2015hvg}. The impact of these time delays is modeled through different scenarios, each with distinct implications for merger rates and characteristics\cite{Boylan-Kolchin:2007bvo,Barausse:2020mdt,Khan:2016vln,khan2012formation}.

\begin{figure}[ht]
    \centering
    \begin{subfigure}[b]{0.49\textwidth}
        \centering
        \includegraphics[height=6.5cm, width=8.5cm]{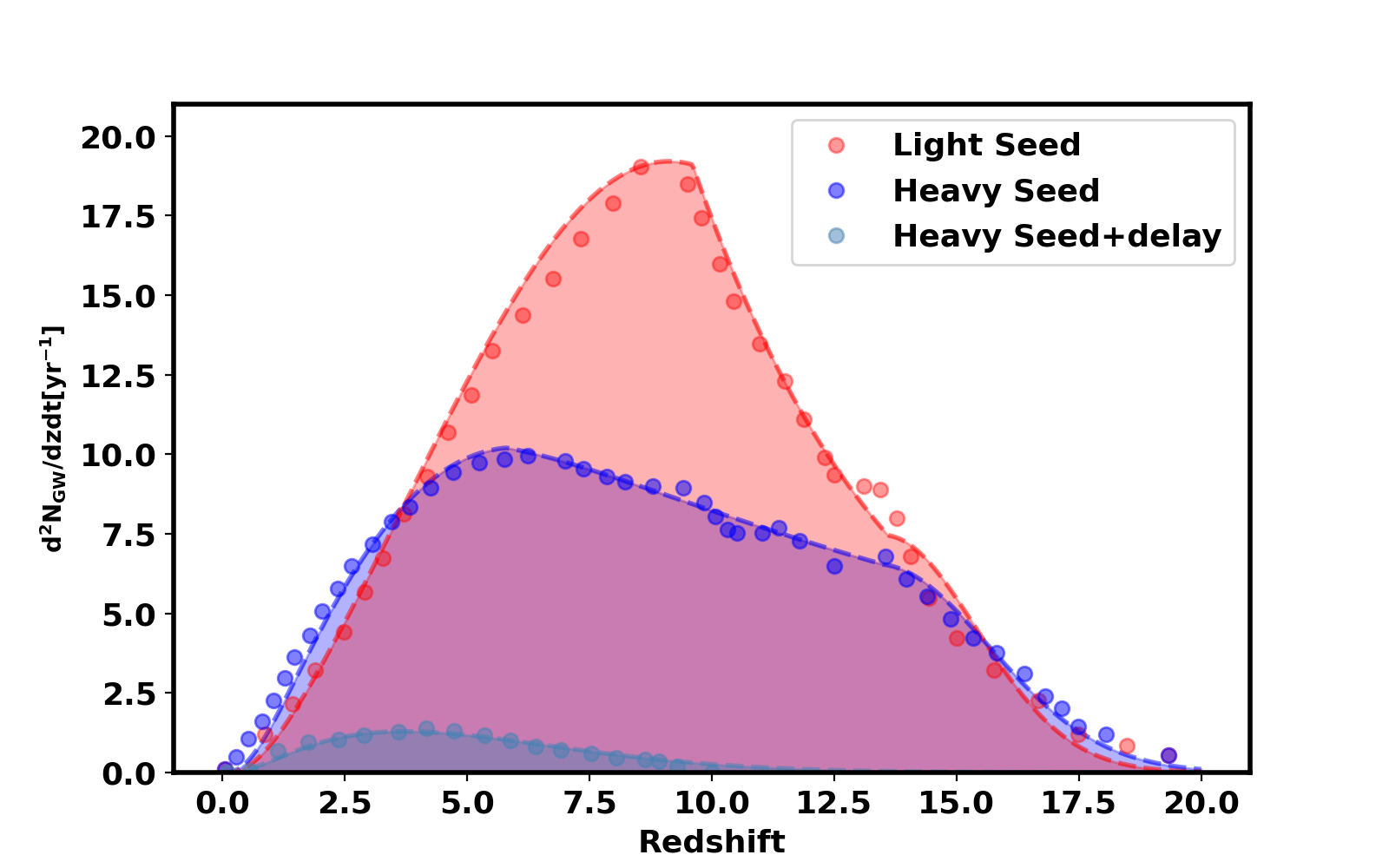}
    \end{subfigure}
    \hfill
    \begin{subfigure}[b]{0.49\textwidth}
        \centering
        \includegraphics[height=6.5cm, width=8.5cm]{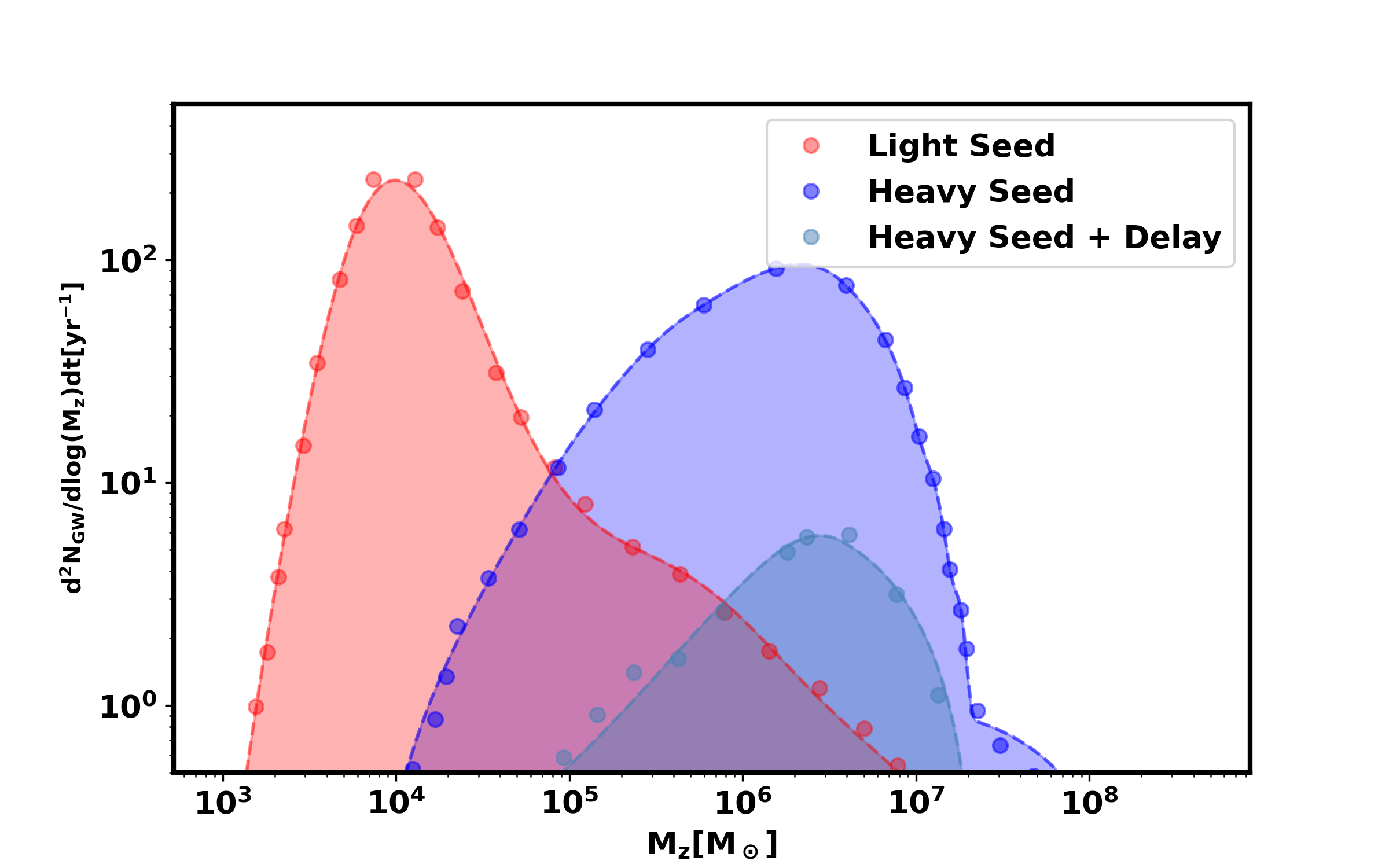}
    \end{subfigure}
    \caption{
        The left plot showcases three merger models for MBBHs. The Light Seed model, featuring light MBH seeds from Light Seed stars, includes delays between MBBHs and galaxy mergers, making it a more conservative scenario. The Heavy Seed+delay model posits heavy MBH seeds from protogalactic disk collapses, also incorporating these delays. Conversely, Heavy Seed excludes such delays, presenting an optimistic upper bound for LISA event rates. All models are fitted with a combination of two Gaussian distributions and a power law function, represented by dashed lines, to illustrate the variation in merger rates across different assumptions. The right plot compares the mass(total mass) models for MBBHs under three scenarios: Light Seed with light MBBHs seeds, Heavy Seed+delay with heavy MBH seeds incorporating merger delays, and Heavy Seed without delays. The models highlight how seed mass and merger timing assumptions influence the predicted MBH mass distribution.
    }
    \label{fig:MergModel}
\end{figure}

All of these factors and their combinations affect the event rate and the mass model of these MBBHs, which are crucial for understanding the range of scenarios detectable by LISA (for more details, see reference \cite{Barausse:2020mdt,Barausse:2012fy}). These elements collectively influence our predictions of the frequency and characteristics of MBBHs merger events. The actual mechanisms of MBH mergers and the corresponding mass models remain largely unknown. The total number of compact binary coalescences per unit redshift is estimated by the equation
\begin{equation}
\mathrm{\frac{dN_{GW}(z)}{dzdt} = \frac{R(z)}{1+z} \frac{dV_c}{dz}},
\label{eq:totno}
\end{equation}
where $\mathrm{\frac{dV_c}{dz}}$ represents the comoving volume element, and $\mathrm{R(z)}$ indicates the merger rate. To capture the wide range of possible merger rate scenarios, we parameterize the merger rate $\mathrm{R(z)}$ using a combination of two Gaussian distributions and a power law, as described by the following equation.

\begin{equation}
\mathrm{R(z) = A}
\begin{cases} 
\mathrm{\exp\left(-\frac{(z - \mu_L)^2}{2 \cdot \sigma_L^2}\right)} & \mathrm{\text{if } z < \mu_L} \\[10pt]
\mathrm{\exp\left(-\alpha \cdot (z - \mu_L)\right)} & \mathrm{\text{if } \mu_L \leq z \leq \mu_R} \\[10pt]
\mathrm{\exp\left(-\frac{(z - \mu_R)^2}{2 \cdot \sigma_R^2}\right) \cdot \exp\left(-\alpha \cdot (\mu_R - \mu_L)\right)} & \mathrm{\text{if } z > \mu_R}
\label{eq:mergerrate}
\end{cases}
\end{equation}
where $\mathrm{A}$ is the overall normalization, $\mathrm{\mu_L}$ is the mean of the left Gaussian distribution, $\mathrm{\sigma_L}$ is the standard deviation of the left Gaussian distribution, $\mathrm{\mu_R}$ is the mean of the right Gaussian distribution, $\mathrm{\sigma_R}$ is the standard deviation of the right Gaussian distribution, and $\mathrm{\alpha}$ is the decay parameter of the exponential function. This approach is motivated by the need to cover all possible variations in the merger rate. Different scenarios can lead to different distributions of mergers, and a flexible parametric form allows us to represent this diversity. The Gaussian distributions help capture localized features in the merger rate, while the power law accounts for the broader trends over redshift. By configuring these components with distinct parameters, we can match a variety of theoretical models and observational constraints. Similarly, the mass model is described using variable parameters that account for different theoretical combinations of the aforementioned factors. For each unique parameter set in the merger and mass models, the predicted number of observable events varies. By integrating Equation \ref{eq:totno}, we calculate the total number of events, allowing us to predict the observable merger events for LISA under different theoretical scenarios. Figure \ref{fig:MergModel} displays the merger rates for three distinct MBBH models using LISA. The Light Seed model involves light MBBH seeds derived from Light Seed stars, considering the temporal delays between MBBHs and galaxy mergers, thereby providing a more conservative estimate. The Heavy Seed+delay model assumes heavy MBBH seeds originating from the collapse of protogalactic disks with similar delays, while Heavy Seed omits these delays, offering an optimistic scenario for LISA event rates.

Each model is fitted using two Gaussian distributions and a power law function, as defined in Equation \ref{eq:mergerrate}, and is depicted with dashed lines in Figure \ref{fig:MergModel}. The dotted points in Figure \ref{fig:MergModel} represent data from semi-analytical simulations, as referenced in\cite{Barausse:2020mdt,Barausse:2012fy}\footnote{\url{https://people.sissa.it/~barausse/catalogs/readme.pdf}}. The solid lines are obtained by fitting these data using Equation \ref{eq:mergerrate}, and all parameter values are listed in Table \ref{tab:ParamValue}. This figure also illustrates the corresponding total redshifted mass $\mathrm{M_z \equiv (1+z)(m_1+m_2)}$ models for these scenarios in terms of the individual masses $\mathrm{m_1}$ and $\mathrm{m_2}$. In addition to the total mass, another crucial parameter for binary systems is the mass ratio ($\mathrm{q \equiv m_1/m_2}$, where $\mathrm{m_1 \leq m_2}$). In this analysis we are interested in measuring any deviation from GR on the cosmological scale. So, we particularly focus on the high matched filtering signal to noise ratio (SNR) events in this analysis so that the uncertainty on any inferred deviation from GR is minimal. So, we consider events above SNR of fifty and only a sub-set of high SNR events are assumed to have EM counterparts. As the events with highest matched filtering SNR are those with mass ratio ($\mathrm{q=1}$) equal to one, we limit our analysis only for $\mathrm{q=1}$ in this analysis. Also, as it is currently unclear on how many LISA sources can have EM counterparts, we obtain forecast on the parameter $\mathcal{F}(z)$ for different scenarios of number of events with EM counterpart in this analysis, ranging from about $4\%$ to $75\%$ events in four years with matched filtering SNR above fifty.

However, it is important to note that LISA binaries can detectable for mass ratios less than one as well and a fraction of them can also have EM counterpart. In this analysis we consider only high SNR events and make a pessimistic choice by not considering EM counterpart to $q<1$ sources. For q within the range of $\mathrm{10^{-3}}$ to $\mathrm{1}$, the probability on $\mathrm{q}$ defined as $\mathrm{P(q)}$ in the range of q values $\mathrm{0.5}$ to $\mathrm{1}$ is always at least 50\% for flat ($\mathrm{P(q)=}$ constant) and for positive power law distributions with $\mathrm{P(q) \propto q^\alpha, \, \alpha>0}$ \cite{Mangiagli:2022niy}. This percentage decreases if we assume a power law with a negative exponent in q. For other $\mathrm{q}$ distributions, we have shown the results in Appendix \ref{sec:appendix}. To select a fraction of sources with the value of mass ratio close to one having EM counterpart, in this analysis we consider scenarios, namely $\mathrm{75\%}$ (fiducial case) and $\mathrm{50\%}$ of the total events after applying the SNR cutoff. We also consider a pessimistic case with only six events detected up to redshift $z=6$ with EM counterpart in four years, which corresponds to a scenario of about $4\%$ events for the Light Seed model and Heavy Seed model, and about $30\%$ events for the Heavy Seed+ delay model.

\begin{table}
\centering
\begin{tabular}{|l|c|c|c|c|c|c|}
\hline
Model & $\mathrm{A}$ & $\mathrm{\mu_L}$  & $\mathrm{\mu_R}$ & $\mathrm{\sigma_L}$  & $\mathrm{\sigma_R}$ & $\mathrm{\alpha}$ \\
\hline
Light Seed & 0.056 & 9.6 & 13.6 & 4.8 & 1.9 & 0.21 \\
\hline
Heavy Seed & 0.028 & 5.8 & 13.5 & 5 & 2.3 & 0.04 \\
\hline
Heavy Seed + Delay & 0.0045 & 1.5 & 1.8 & 10.5 & 4.5 & 0.02 \\
\hline
\end{tabular}
\caption{Parameter values for different parameters in the models used in the merger rate $\mathrm{R(z)}$ calculations. The parameters include amplitude $\mathrm{A}$, mean $\mathrm{\mu_L}$ and standard deviation $\mathrm{\sigma_L}$ for left Gaussian, mean $\mathrm{\mu_R}$ and standard deviation $\mathrm{\sigma_R}$ for right Gaussian, and decay parameter $\mathrm{\alpha}$.}
\label{tab:ParamValue}
\end{table}

Galaxy simulations to forecast the EM counterparts for massive binary mergers typically assume a well-understood evolution of galaxies and star formation rates, calibrated from large-scale surveys. However, simulating the formation and distribution of galaxies, especially at high redshifts, comes with significant uncertainties\cite{2017ARA&A..55...59N,2023IAUS..373..283N,Sales:2022ich}. One of the major challenges arises from the assumptions about gas accretion, feedback processes, and the precise merger history, which can vary significantly across different simulation models. Recent observations from the JWST have further complicated our understanding of the high-redshift universe. JWST has revealed a higher abundance of massive galaxies and earlier star formation than previously predicted, challenging existing models of galaxy formation and evolution. JWST has also detected a variety of galaxies with different dust and gas content, which directly affects the probability of producing an EM counterpart. Galaxies rich in gas are more likely to produce bright EM counterparts during binary mergers. Given these uncertainties, we assume that approximately 75\% and 50\% of massive binary mergers may have observable EM counterparts, depending on their redshift and environment. We have also shown the results for scenarios with $25\%$ and $10\%$ bright sirens in the appendix \ref{sec:ErrorBudget}. This assumption reflects the current understanding of galaxy formation processes and the likelihood that such mergers would occur in gas-rich environments, which are more conducive to generating EM radiation. While these fractions are estimates, they align with predictions from theoretical models and observational data, with the emerging JWST results suggesting a broader range of environments that could host potential EM counterparts. The models considered here are related to different theoretical approaches to MBBHs formation and evolution, significantly affecting the predicted total mass distributions and influencing the detectability of merger events by LISA. The plots effectively demonstrate how variations in initial assumptions about MBBHs seed formation and merger delays impact the mass spectrum observable by future missions. It is important to that all of these events cannot be detected due to the limitations imposed by the sensitivity of detectors. The calculation of the matched filtering SNR is essential to determine which events are detectable.

\subsection{Signal to Noise Ratio (SNR) for LISA source} The matched filtering SNR, denoted as $\mathrm{\rho}$, is crucial for detecting GW events as it quantifies the GW signal's strength relative to the background noise. An event is deemed detectable if its SNR surpasses a predefined threshold, $\mathrm{\rho_{TH}}$. The optimal SNR for a GW emitted by a binary system is defined by the equation \cite{Sathyaprakash:1991mt,Cutler:1994ys,Balasubramanian:1995bm}
\begin{equation}
    \mathrm{\rho^2 = 4 \int_{f_{\text{min}}}^{f_{\text{max}}} \frac{|\tilde{h}(f)|^2}{S_n(f)} \, df},
\end{equation}
where $\mathrm{\tilde{h}(f)}$ represents the Fourier transform of the GWs strain, while $\mathrm{S_n(f)}$ denotes the power spectral density of the detector noise. 
\begin{figure}
\centering
\includegraphics[height=6.5cm, width=14cm]{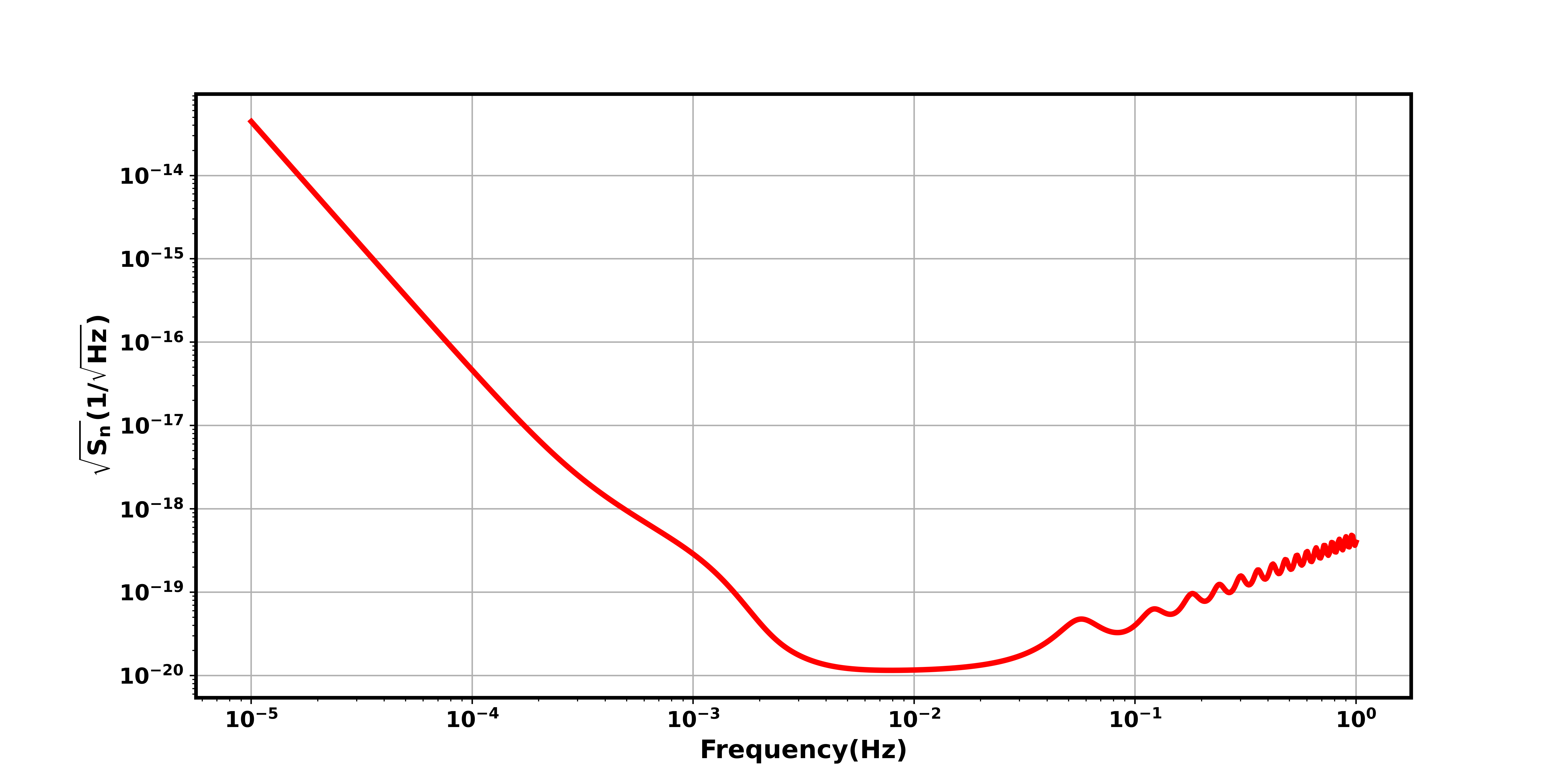}
\caption{Noise curve for the LISA showing the square root of the power spectral density, $\mathrm{\sqrt{S_n}}$, as a function of frequency. The sensitivity is depicted in units of $\mathrm{1/\sqrt{Hz}}$.}
\vspace{-0.5cm}
\label{fig:Noise}
\end{figure}
For $\mathrm{S_n(f)}$, we adopt the semi-analytical forms from \cite{Babak:2021mhe}, as shown in Figure\ref{fig:Noise}, and $\mathrm{f_{min}}$ and $\mathrm{f_{max}}$ are the boundaries for the frequency integral. In our study, for the computation of the SNR, we set $\mathrm{f_{max}}$ to the frequency of the innermost stable circular orbit ($\mathrm{f_{ISCO}}$) and $\mathrm{f_{min}}$ to the frequency one year prior to $\mathrm{f_{ISCO}}$, considering our observation period is one year. We do not include the Galactic background noise in the power spectral density. The Galactic background can have an impact on the overall noise, which in turn affects the number of detectable events.

Specifically, $\mathrm{\tilde{h}(f)}$ is expressed as
\begin{equation}
    \mathrm{\tilde{h}(f) = h(f)Q},
\end{equation}
with $\mathrm{h(f)}$ denoting the strain amplitude
\begin{equation}
\mathrm{h(f)=\sqrt{\frac{5\eta}{24}} \frac{(GM_{\text{chirp}})^{5/6}}{D_L\pi^{2/3}c^{3/2}}f^{-7/6}e^{-i \psi(f)}},  
\end{equation}
where $\mathrm{M_{chirp}}$ is the chirp mass, $\mathrm{\eta}$ is the mass ratio, $\mathrm{D_L}$ is the luminosity distance to the source, $\mathrm{c}$ is the speed of light, and $\mathrm{\psi(f)}$ is the phase of the waveform \cite{Ajith:2007kx}. The factor $\mathrm{Q}$ depends on the inclination angle and the detection antenna functions $\mathrm{F_+}$ and $\mathrm{F_\times}$
\begin{equation}
    \mathrm{Q = \left[F_+ \left(\frac{1+\cos^2(\iota)}{2}\right) + \iota F_\times \cos(\iota)\right]}.
\end{equation}
The antenna functions are defined by\cite{Babak:2021mhe}
\begin{align}
    \mathrm{F_{+}} &= \mathrm{\frac{1}{2}(1+\cos^2\theta)\cos2\phi \cos2\psi - \cos\theta \sin2\phi \sin2\psi}, \\
    \mathrm{F_{\times}} &= \mathrm{\frac{1}{2}(1+\cos^2\theta)\cos2\phi \sin2\psi + \cos\theta \sin2\phi \cos2\psi},
\end{align}
where $\mathrm{\theta}$ and $\mathrm{\phi}$ determine the source's location in the sky, and $\psi$ relates to the orientation of the binary system with respect to the detector \cite{Finn:1992xs}. Consequently, the SNR formula becomes 
\begin{equation}
    \mathrm{\rho = \frac{\Theta}{4}\left[4\int_{f_{\text{min}}}^{f_{\text{max}}} \frac{h(f)^2}{S_n(f)} df\right]^{1/2}},
    \label{eq:SNRcalc}
\end{equation}
where $\mathrm{\Theta^2 \equiv 4 \left(F_{+}^2(1+\cos^2i)^2 + 4F_{\times}^2\cos^2i\right)}$. According to  \cite{Finn:1995ah}, averaging over many binaries, the inclination angle, and sky positions, $\mathrm{\Theta}$ follows the distribution
\begin{equation}
\mathrm{P(\Theta)} = 
\begin{cases}
\mathrm{5\Theta(4-\Theta)^3/256} & \text{if } \mathrm{0<\Theta<4},\\
\mathrm{0} & \text{otherwise}.
\end{cases}
\end{equation}

\begin{figure}
\centering
\includegraphics[height=6.5cm, width=15cm]{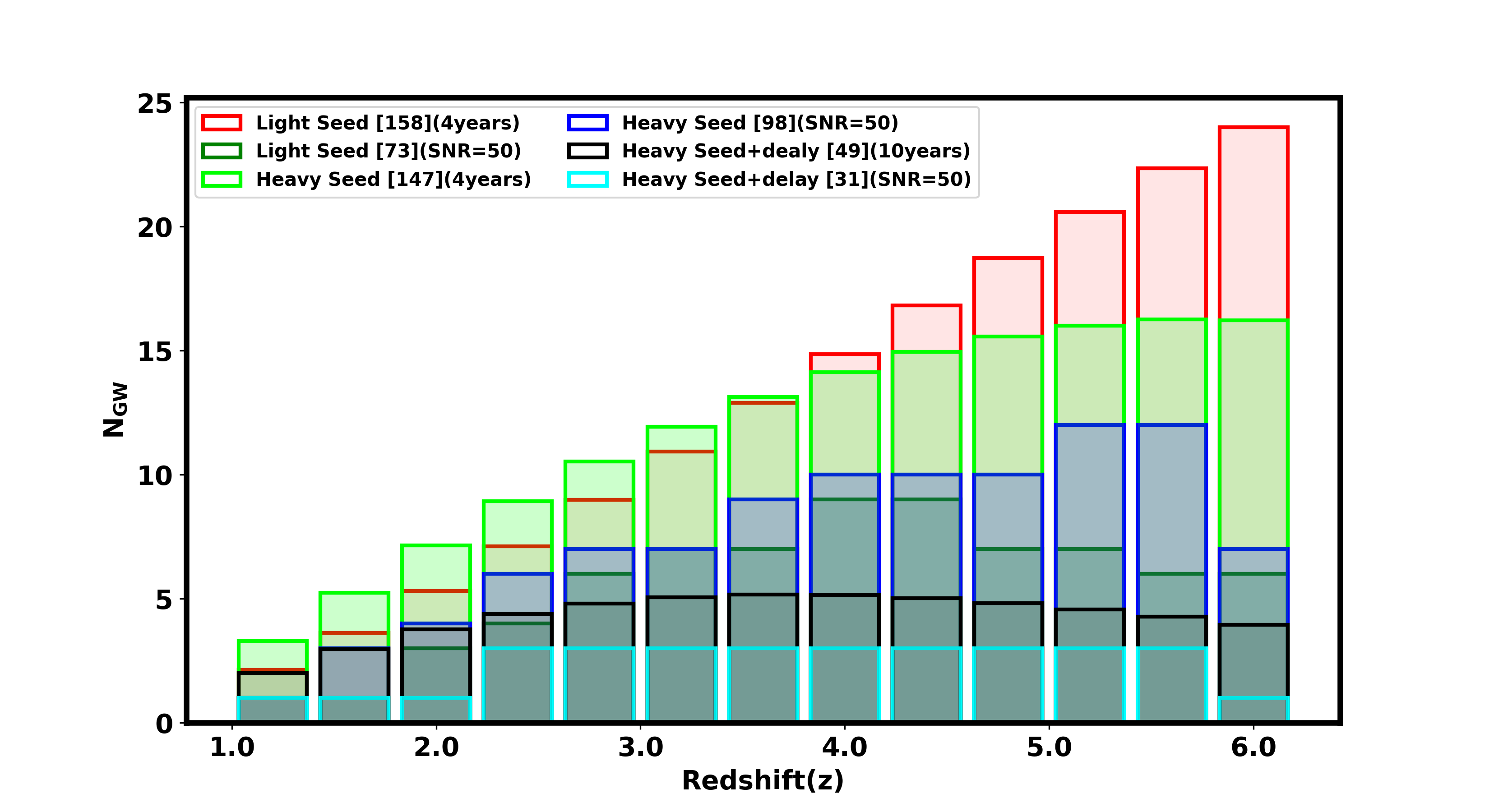}
\caption{The figure presents the total number of injected GW events and the number of detectable GW events as a function of redshift for three different models: Light Seed, Heavy Seed, and Heavy Seed+delay. The total observation periods are indicated in parentheses next to the total number of injected events. For the detectable events, the threshold SNR is mentioned in parentheses next to the number of detectable events, assuming that EM counterparts are detected for $\mathrm{75\%}$ of the events.}

\vspace{-0.5cm}
\label{fig:EventCount}
\end{figure}

\begin{figure}[ht]
    \centering
    \includegraphics[height=6.5cm, width=17cm]{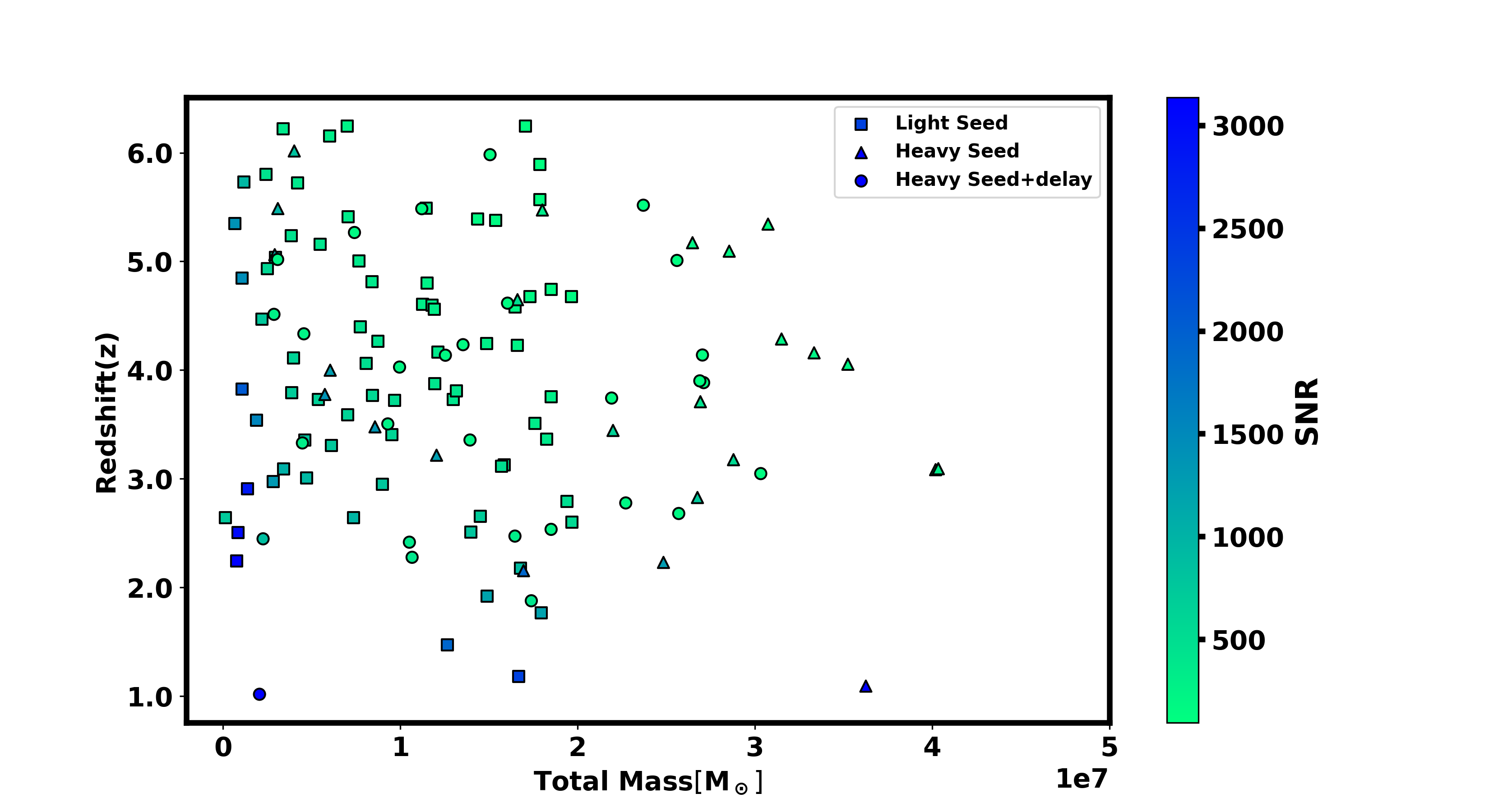}
    \caption{In this figure, we present a scatter plot illustrating the total mass of the two components of the binary system against the redshift of the selected events, with a threshold SNR of 50. The plot includes data for three different models: Light Seed, Heavy Seed, and Heavy Seed+delay, represented by square, triangle, and circular markers, respectively. The observations span 4 years for the Light Seed and Heavy Seed models and 10 years for the Heavy Seed+delay model, assuming that EM counterparts are detected for $\mathrm{75\%}$ of the events. The color of each point represents the SNR of the event, providing an additional dimension of information.}
    \label{fig:SNRandMASS}
\vspace{-0.5cm}
\end{figure}

In this study, we investigate various merger and mass models, recognizing that our understanding of these models is still incomplete. By examining different scenarios, we demonstrate how different assumptions about the formation and evolution of MBBHs impact the measurements of $\mathrm{\mathcal{F}(z)}$ using LISA sources. To estimate the total number of events, we integrate using Equation \ref{eq:totno} and determine the number of merging events within each redshift bin, selecting a bin size of $\mathrm{\Delta z = 0.4}$. Figure \ref{fig:EventCount} displays the number of events binned by redshift for the Light Seed, Heavy Seed, and Heavy Seed + Delay models. The total number of events for each model, along with the observation years, is summarized in Table \ref{tab:EventSummary}. For each event, we use the inverse transform method to derive the total mass of the binary system and the parameter $\mathrm{\Theta}$ from their probability distributions. We then generate an equal-mass binary system and calculate the SNR using Equation \ref{eq:SNRcalc}. 

The events with an SNR exceeding the threshold are selected. Since we only consider a mass ratio $\mathrm{q}$ of 1 when generating the binary sources, we retain $\mathrm{75\%}$ of the events after the SNR cutoff. The number of events selected if only $\mathrm{50\%}$ are retained is also noted and listed in Table \ref{tab:EventSummary}. In Sec. \ref{sec:ErrorBudget}, we have also shown the results for different fractions of bright standard sirens detectable from LISA. For comparison, in \cite{Mangiagli:2022niy}, it was found that, for a mission duration of 5 years with an 80\% duty cycle up to a redshift of 10 and an SNR cutoff of 10, there were a few hundred events for both light seed and heavy seed scenarios, and about 30 events for the heavy seed with delay scenario. Here, we consider a more pessimistic choice for SNR and redshift when selecting EM bright sources, as we focus only on high SNR golden events for testing GR (see Sec. \ref{sec:result} and appendix \ref{sec:ErrorBudget}). The detection probability for the Light Seed model is lower compared to the other two models because its mass distribution peaks at much lower values, resulting in reduced signal strength and, consequently, lower SNR. Figure \ref{fig:SNRandMASS} illustrates the total mass of the two components of the binary system as a function of the redshift of the selected events, with an SNR threshold of fifty.

\begin{table}
\centering
\begin{tabular}{|l|c|c|c|c|c|}
\hline
Model & $T_{obs}$ (years) & Total Events & Selected (75\%) & Selected (50\%) & 1 Event Detectability \\
\hline
LS & 4 & 158 & 73 & 49 & 6.12\% \\
\hline
HS & 4 & 147 & 98 & 66 & 4.54\% \\
\hline
HSD & 10 & 49 & 31 & 21 & 14.28\% \\
\hline
\end{tabular}
\caption{Number of total and detected events for different merger models Light Seed (LS), Heavy Seed (HS), and Heavy Seed+Delay (HSD), with a threshold SNR of 50. The table includes results based on 75\% and 50\% selection criteria, as well as the percentage of detectability of a single event at each from $z=1$ to $z=6$ (1 event detectability).}
\label{tab:EventSummary}
\end{table}

\subsection{LISA Source Parameter Estimation}
\label{sec:ParamEst}

LISA will consist of a constellation of three spacecraft forming a giant equilateral triangle in space, each containing free-falling test masses. Lasers measure the minute changes in distance between these masses, changes induced by passing gravitational waves. However, a significant challenge in detecting these waves is the overwhelming noise from laser frequency fluctuations \cite{2017arXiv170200786A}. To combat this, LISA uses Time-Delayed Interferometry (TDI), a sophisticated data processing technique that combines the laser measurements in a way that reduces noise by several orders of magnitude, making the subtle signals of gravitational waves detectable. The parameter estimation process in LISA involves several complex steps. First, the raw data from the TDI observables specifically the second-generation observables X, Y, and Z, which accommodate the fact that the spacecraft formation allows for non rigid arm lengths are transformed into the A, E, and T channels  \cite{Tinto:2000qij,Tinto:2003vj,Estabrook:2000ef,Vallisneri:2005ji,Tinto:2004wu,Tiwari:2023mzf,Prince:2002hp}. These channels are designed to be approximately uncorrelated in terms of their noise properties, improving the clarity and reliability of the data for further analysis.

\begin{figure}[ht]
\centering
\includegraphics[height=10.0cm, width=15cm]{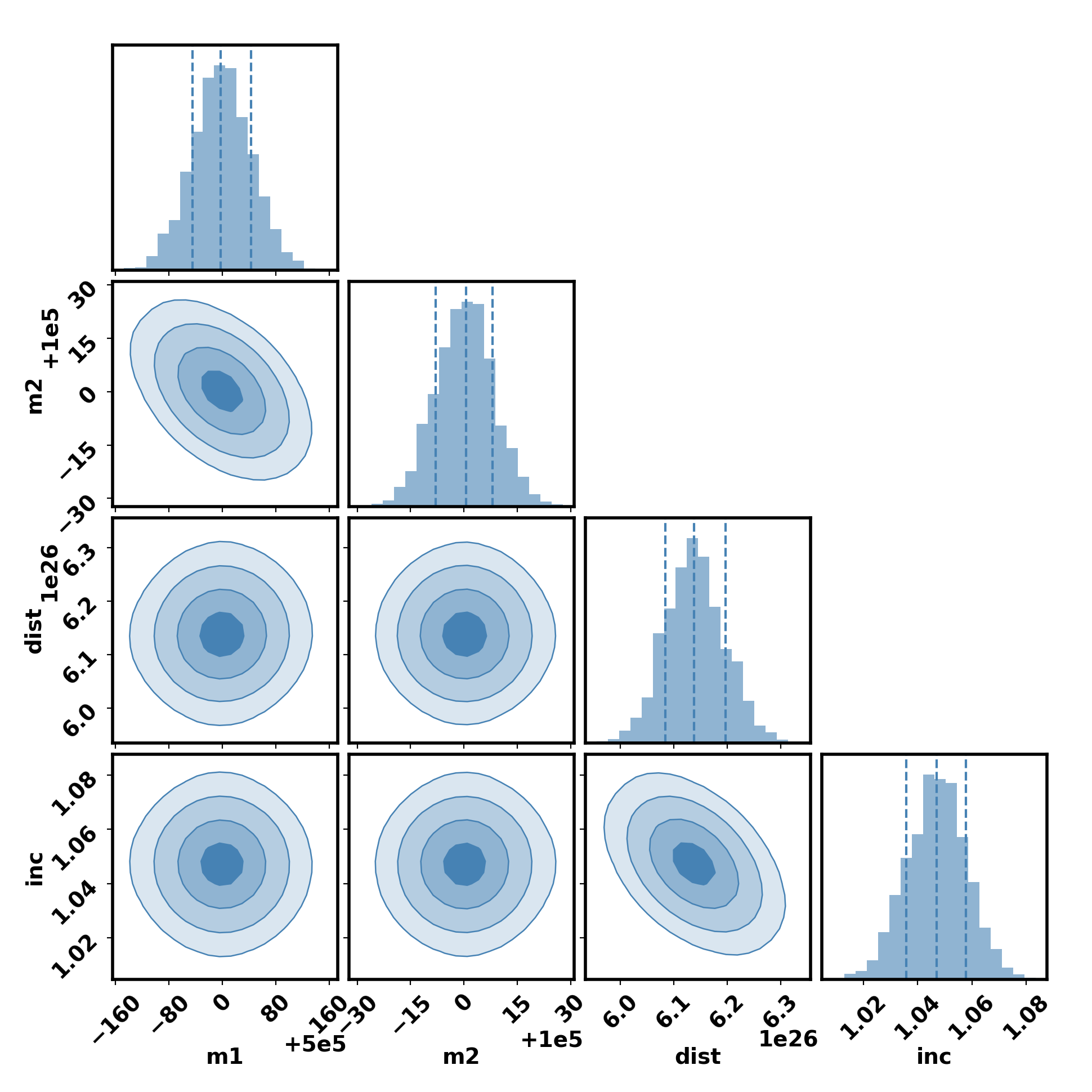}
\caption{This plot presents the estimated values and uncertainties for the component masses ($\mathrm{m_1}$ and $\mathrm{m_2}$), luminosity distance, and inclination angle for LISA source at a redshift of $\mathrm{z=2.4}$. These distributions highlight the precision achievable with LISA's observational capabilities, underscoring the potential for profound cosmological insights into the properties of binary systems in the early universe. The injected values are as follows: black hole masses $\mathrm{m_1 = 5 \times 10^5 M_\odot}$, $\mathrm{m_2 = 1 \times 10^5 M_\odot}$, luminosity distance (denoted by dist)  $\mathrm{6.137 \times 10^{26}}$ meters (19.87 Gpc), and inclination angle (denoted by inc) of 1.05 radians.}
\vspace{-0.5cm}
\label{fig:ParamEst}
\end{figure}

\begin{figure}
\centering
\includegraphics[height=6.5cm, width=13cm]{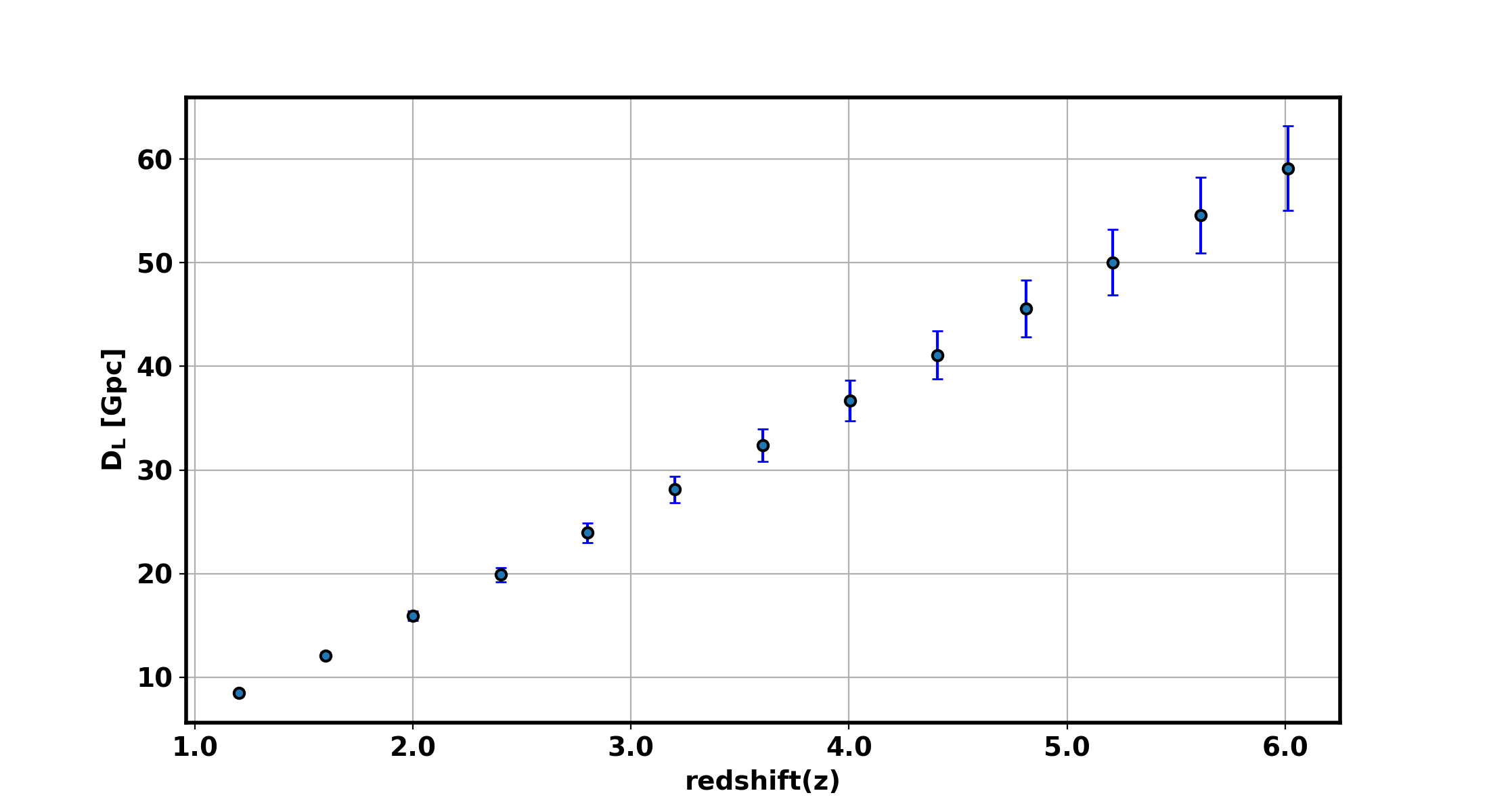}
\caption{This figure illustrates the luminosity distance errors as a function of redshift for LISA sources, enhanced to include detailed error analyses. Each data point represents the measured luminosity distance along with the total uncertainty, derived from two specific sources of errors: GW source parameter estimation errors and WL errors. The error bars correspond to $1$-$\sigma$ uncertainty. The figure highlights the precision and challenges LISA faces in distance estimation across various redshifts and shows how the precision of measurements varies with increasing redshift.}
\vspace{-0.5cm}
\label{fig:disterr}
\end{figure}

To study the GW events detected by the LISA, it is essential to estimate the parameters of these sources. The process of parameter estimation for LISA sources is quite different from that used for ground-based observatories like LIGO, Virgo, and KAGRA \cite{Hoy:2023ndx}. For our analysis of LISA sources, we employ the \texttt{BBHx} Python package\cite{Katz:2020hku} , which is specifically designed for this purpose. Our methodology involves generating waveforms using \texttt{BBHx} \cite{Katz:2020hku}, which incorporates the \texttt{IMRPhenomD} \cite{Klein:2015hvg,Husa:2015iqa} waveform approximation model. We obtain the waveform for a one-year observation period of the LISA sources. Subsequently, we use \texttt{BBHx} \cite{Katz:2020hku} to calculate the likelihood of the generated waveforms. For the estimation of the posterior distributions of key parameters for this study specifically, the masses of the black holes ($\mathrm{m_1}$ and $\mathrm{m_2}$), the luminosity distance ($\mathrm{D_L}$), and the inclination angle ($\mathrm{i}$) we constrain the priors of all other parameters to delta functions. Finally, for the parameter estimation, we employ the \texttt{Dynesty} \cite{speagle2020dynesty} sampler. Among these, the posterior of $\mathrm{D_L}$ is particularly crucial for our study, as it will be used for the inference of $\mathrm{\mathcal{F}(z)}$.

The accuracy of luminosity distance measurements in GW astronomy is influenced not only by the sensitivity and configuration of the detector but also significantly by weak lensing, especially for sources detected at high redshifts by LISA \cite{Holz:2005df}. Gravitational lensing affects GWs similarly to EM radiation. For GW events expected from large redshifts ($\mathrm{z > 1}$), weak lensing is a common occurrence, along with the occasional strongly-lensed sources. A lens with magnification $\mu$ modifies the observed luminosity distance to $\mathrm{\frac{D_L}{\sqrt{\mu}}}$, introducing a systematic error $\mathrm{\Delta D_L / D_L = 1 - \frac{1}{\sqrt{\mu}}}$ \cite{Mpetha:2022xqo}. This lensing error, when convolved with the expected magnification distribution $\mathrm{p(\mu)}$ from a standard $\mathrm{\Lambda}$CDM model, significantly influences the overall measurement accuracy. In this study, we utilize the following model to estimate the error due to weak lensing \cite{Hirata:2010ba}
\begin{equation}
    \mathrm{\frac{\sigma_{WL}}{D_L} = \frac{0.096}{2} \left(\frac{1 - (1 + z)^{-0.62}}{0.62}\right)^{2.36}}.
\end{equation}
The total uncertainty in the measured luminosity distance, therefore, is expressed as
\begin{equation}
    \mathrm{\sigma_{D_L}^2 = \sigma_{GW}^2 + \sigma_{WL}^2},
\end{equation}
where $\mathrm{\sigma_{GW}}$ is the uncertainty derived from the detector's setup during parameter estimation, and $\mathrm{\sigma_{WL}}$ accounts for the lensing effects. This comprehensive approach ensures a more accurate understanding of the intrinsic properties of GW sources.

The figure \ref{fig:ParamEst} shows the estimated values and uncertainties for component masses, luminosity distance, and inclination angle of LISA source at a redshift of z=2.4. This highlights the potential of LISA to offer precise insights into the cosmological properties of binary systems from the early universe. The figure \ref{fig:disterr} presents the luminosity distance errors as a function of redshift for LISA sources, showcasing the variability and precision of measurements across different redshifts. Each point in the plot represents the measured luminosity distance at various redshifts, with total uncertainty, derived from two specific sources of errors: GW source parameter estimation error and WL error. This figure underscores the advanced capabilities of LISA in probing the universe's most distant reaches, illustrating the impact of different sources of uncertainty on measurement accuracy, and paving the way for profound cosmological discoveries.

\section{Identification of EM Counterparts for LISA Sources}
\label{sec:EMcounterparts}

The EM counterpart detection for LISA sources involves a sophisticated interplay of various observational techniques and instruments across multiple wavelengths. One of the primary targets for such detections is the optical emission from AGN, which can be observed using the Vera Rubin Observatory \cite{Merloni:2008hx,Shen:2003sda,Teukolsky:2014vca,Shen:2020obl}. The Vera Rubin Observatory, with its powerful 8.4-meter mirror, is capable of capturing images in the u, g, r, i, z, and y bands \cite{LSST:2008ijt}. There are two main strategies for utilizing the Rubin Observatory in the search for LISA counterparts. The first involves sifting through archival data from the Vera Rubin Observatoryto identify potential modulations caused by the proper motion of binaries prior to their merger. The second strategy is to employ Target of Opportunity (ToO) observations, which can be triggered when a LISA event is detected, allowing astronomers to quickly point the telescope and capture fresh data. The depth of the Vera Rubin Observatorysurvey, expected to reach an apparent magnitude of around 27.5 after ten years of operations, sets the detection threshold for these observations \cite{Shen:2020obl,willmer2018absolute,Laigle:2019iws}.

To effectively detect EM radiation from merging MBBHs, it is essential to use radio and optical telescopes to target GWs sources preemptively. For selecting potential candidates that may emit EM counterparts during the inspiral phase of MBBHs, a conservative approach involves choosing GWs events with a SNR of 8 or higher and a sky localization accuracy within 10 square degrees, matching the field of view of the Vera Rubin Observatory\cite{Jin:2023sfc,Mangiagli:2022niy,Tamanini:2016zlh,LISACosmologyWorkingGroup:2019mwx}. For this analysis we have chosen only events with SNR greater than 50, as a result the sky localization error will be better and EM counterparts can be easier to identify.

Among LISA-detected events meeting these criteria, we assume that EM emissions at merger consist of optical flares driven by accretion and radio flares and jets, as indicated by general-relativistic simulations of MBH mergers in external magnetic fields. The semi-analytical simulations within MBBHs catalogs provide detailed data on mass in stars, nuclear gas, and other parameters, which are essential for estimating the magnitude of each merger's EM emission in both optical and radio bands \cite{Palenzuela:2010nf,Gold:2014dta,Kelly:2017xck,Cattorini:2021elw}. The future detection capabilities of EM facilities like Vera Rubin Observatory \cite{abell2009lsst}, the Square Kilometre Array (SKA) \cite{rawlings2011square}, and the Extremely Large Telescope (ELT)\footnote{\url{https://www.eso.org/sci/facilities/eelt/}} can be assessed by determining the expected number of detectable counterparts. While the sky localization region may contain multiple EM transients, the true EM counterpart is assumed to be efficiently identified using methods such as timing, spectrum analysis, and host galaxy information, similar to prior studies. This precise identification allows for the host galaxy to be pinpointed and the GWs parameters to be refined by fixing the sky localization angles, thereby enhancing the accuracy of the measurements.

In addition to optical observations, radio emissions, particularly those emanating from jets, provide another promising avenue for detecting EM counterparts of LISA sources. The SKA, expected to be the world's largest radio telescope, will play a pivotal role in this effort. Significant radio emissions can be produced by the interaction between the plasma surrounding a merging binary and the magnetic fields. These emissions can manifest as flares due to the twisting of magnetic field lines or as jets driven by mechanisms like the Blandford-Znajeck effect  \cite{Okamoto:2019zgf}. The SKA can be employed in ToO observations to capture these radio emissions. Depending on the beaming and the orientation of the jets, the radio luminosity can vary significantly. The ability of SKA to detect these emissions is contingent upon the observed flux surpassing the instrument’s sensitivity threshold, which is set at 1 micro-Jansky for a typical frequency of 1.7 GHz \cite{Meier:2000wk,Yuan:2021jjt,Cohen:2006rt,Moesta:2011bn,Gultekin:2019dtc}.

In the near-infrared spectrum, the ELT is anticipated to be a vital tool for redshift measurements of host galaxies. ELT’s MICADO spectrograph will enable the precise measurement of redshifts in the IYJHK bands. The detection threshold for ELT’s spectroscopy is set at an apparent magnitude of 27.2, while for imaging, it extends to 31.3. For galaxies brighter than 27.2, the redshift measurement error is expected to be extremely small, allowing for highly accurate determinations. For fainter galaxies, between magnitudes 27.2 and 31.3, the precision of redshift measurements varies depending on the actual redshift of the source. High-redshift sources, those beyond a redshift of $\mathrm{z=5}$, can be identified by the Lyman-alpha break with a redshift error of around 0.2. For sources with redshifts between 0.5 and 5, the Balmer break can be used for redshift identification; however, the absence of observations in the optical and UV parts of the spectrum increases the redshift error to about 0.5 \cite{davies2010micado,dunlop2012observing,spergel2015wide,Gardner:2006ky}. Collectively, these strategies and instruments provide a comprehensive framework for the detection and study of the EM counterparts of LISA sources. By combining data across optical, radio, X-ray, and near-infrared wavelengths, a multi-faceted understanding of the processes and environments surrounding merging MBBHs and other GW sources can be gained. This multi-wavelength approach is crucial for the advancement of our knowledge of the Universe and for maximizing the revolutionary capabilities of the LISA mission.

\section{Baryon Acoustic Oscillation Scale from galaxy power spectrum}
\label{sec:BAO}

BAO are patterns resembling wrinkles in the distribution density of galaxies across the universe, originating from acoustic density waves in the early universe's primordial plasma\cite{peebles1973statistical, crocce2011modelling}. These waves were propelled by the interplay between radiation pressure and gravitational forces. As the universe expanded and cooled, leading to the decoupling of photons from baryons, a characteristic scale the sound horizon at the drag epoch ($\mathrm{z_d}$) denoted by $\mathrm{r_s}$ left a discernible imprint on the galaxy distribution\cite{peebles1970primeval, sunyaev1970small, bond1987statistics, hu2002cosmic, blake2003probing}.

\subsection{Inference of BAO from galaxy two-point angular correlation function} The imprint manifests as a distinct peak within the two-point angular correlation function (2PACF) of galaxies, $\mathrm{w(\theta)}$, which offers a method to measure the universe's expansion rate across its history. To detect the BAO feature within $\mathrm{w(\theta)}$, it is modeled using a combination of a power-law and a Gaussian component, represented by the following equation\cite{xu20122, carvalho2016baryon, alam2017clustering}
\begin{equation}
\mathrm{w(\theta, z) = a_1 + a_2\theta^k + a_3\exp\left(-\frac{(\theta - \theta_{\text{FIT}}(z))^2}{\sigma_{\theta}^2}\right)},
\label{eq:fit}
\end{equation}
where $a_i$ are parameters representing the fit, $\mathrm{\sigma_{\theta}}$ is the width of the BAO peak, and $\mathrm{\theta_{FIT}}$ indicates the angular position of the BAO feature. This method offers a robust standard ruler that enables independent estimates of the angular diameter distance, $\mathrm{D_a(z)}$, further elucidating the dynamics of the universe's expansion\cite{peebles1974statistical, davis1983survey, hewett1982estimation, hamilton1993toward, landy1993bias}.

\begin{figure}
\centering
\includegraphics[height=6.5cm, width=13cm]{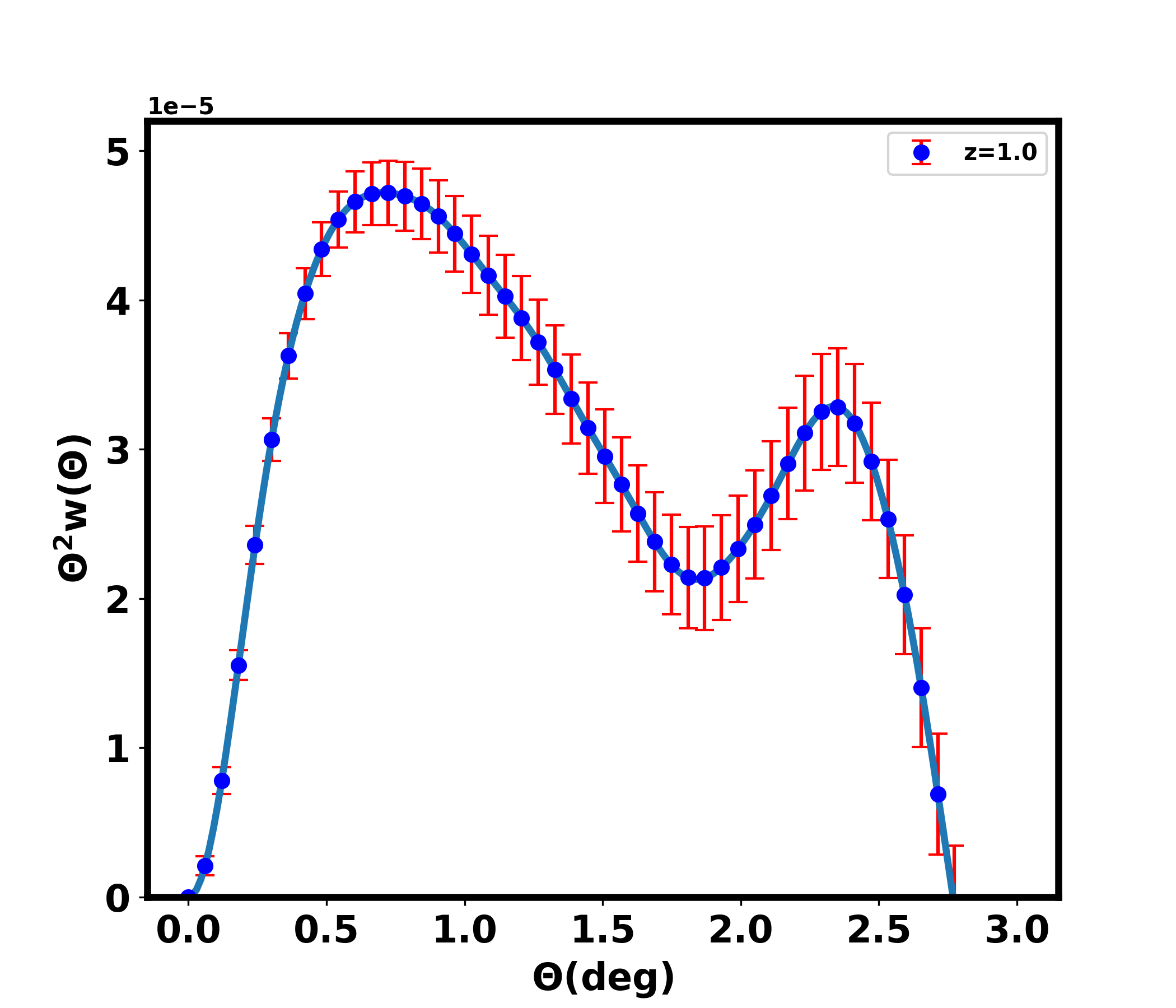}
\caption{This graph displays the relationship of the 2PACF ($\mathrm{\theta^2w(\theta)}$) as it varies with $\mathrm{\theta}$ at a particular redshift, z=1.0, derived from the non linear matter power spectrum. A notable peak appears near $\mathrm{\theta \sim 2.5}$, which signifies the detection of the characteristic BAO scale, denoted as $\mathrm{\theta_{BAO}}$.}
\label{fig:bao}
\vspace{-0.5cm}
\end{figure}

\subsection{Modeling the two-point angular correlation function}
The theoretical representation of the 2PACF, $\mathrm{w(\theta)}$, is formulated as
\begin{equation}
\mathrm{w(\theta) = \sum_{l \geq 0} \left(\frac{2l + 1}{4\pi}\right) P_l(\cos(\theta)) C_l},
\end{equation}
where $\mathrm{P_l}$ denotes the Legendre polynomial of the $\mathrm{l^{th}}$ order, and $\mathrm{C_l}$ represents the angular power spectrum. The latter is derived from the three-dimensional matter power spectrum $\mathrm{\mathcal{P}(k)}$ through the integral\cite{peebles1973statistical, crocce2011modelling}
\begin{equation}
\mathrm{C_l = \frac{1}{2\pi^2} \int 4\pi k^2 \mathcal{P}(k) \psi_l^2(k) e^{-k^2/k_{eff}^2}},
\end{equation}
where $\mathrm{\psi_l(k)}$ is defined as:
\begin{equation}
\mathrm{\psi_l(k) = \int dz \phi(z) j_l(kr(z))},
\end{equation}
incorporating $\mathrm{\phi(z)}$, the galaxy selection function, and $\mathrm{j_l}$, the spherical Bessel function of the $\mathrm{l^{th}}$ order, with $\mathrm{r(z)}$ being the comoving distance. An exponential damping factor $\mathrm{e^{-k^2/k_{eff}^2}}$ is included to enhance the integration's convergence in calculating the angular power spectrum, where $\mathrm{1/k_{eff} = 3 \text{ Mpc/h}}$ is consistently used across all calculations, proven to have negligible effects on the scales under consideration\cite{anderson2012clustering}. For deriving the matter power spectrum, we employ the \texttt{CAMB}\cite{lewis2011camb} module, configuring the galaxy selection function as a normalized Gaussian distribution. We set the standard deviation of this distribution to $\mathrm{0.03(1+z)\%}$ of its mean value, aligning it with the anticipated photo-z error from the Vera Rubin Observatory \cite{abell2009lsst, ivezic2019lsst, mandelbaum2018lsst} .

\begin{figure}
\centering
\includegraphics[height=6.5cm, width=17cm]{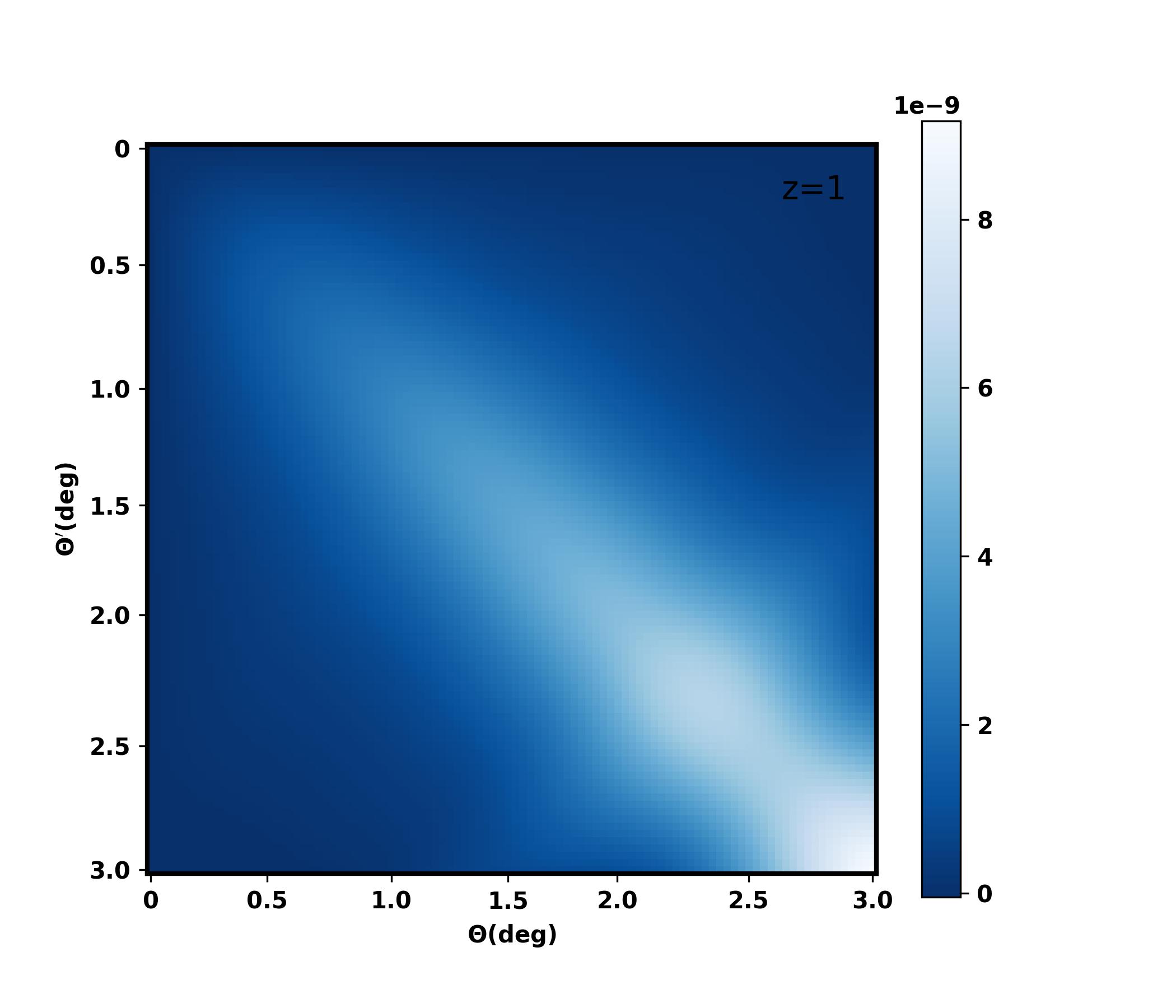}
\caption{This graph presents the covariance matrix ($\mathrm{Cov_{\theta\theta'}}$), computed from the nonlinear matter power spectrum, plotted against $\mathrm{\theta}$ and $\mathrm{\theta'}$ at a specific redshift, $\mathrm{z=1.0}$. The plot clearly shows that the values of the covariance matrix rise as both $\mathrm{\theta}$ and $\mathrm{\theta'}$ increase, suggesting a greater error at larger angular scales.}
\label{fig:covmatplot}
\vspace{-0.5cm}
\end{figure}

Figure \ref{fig:bao} showcases the variation of $\mathrm{\theta^2w(\theta)}$ with angular separation at a fixed redshift $\mathrm{z=1.0}$. A marked peak at approximately $\mathrm{\theta \sim 2.5}$ reveals the BAO scale, central to our analysis. Furthermore, the covariance of $\mathrm{w(\theta)}$, expressed as $\mathrm{Cov_{\theta\theta'}}$, is computed as:

\begin{equation}
\mathrm{Cov_{\theta\theta'} = \frac{2}{f_{sky}} \sum_{l \geq 0} \frac{2l + 1}{(4\pi)^2} P_l(\cos(\theta)) P_l(\cos(\theta')) \left(C_l + \frac{1}{n}\right)^2},
\end{equation}
where $\mathrm{1/n}$ signifies the shot noise linked to the galaxy's number density per steradian, and $\mathrm{f_{sky}}$ is the fraction of the sky surveyed\cite{cabre2007error, dodelson2020modern}. A depiction (Figure \ref{fig:covmatplot}) outlines how the covariance matrix evolves with angular separations at a given redshift(z=1), indicating a growing correlation as $\mathrm{\theta}$ and $\mathrm{\theta'}$ increase. We have included the complete covariance matrix in the analysis, as there are sufficient correlations between different scales.

\begin{figure}
\centering
\includegraphics[height=6.5cm, width=15cm]{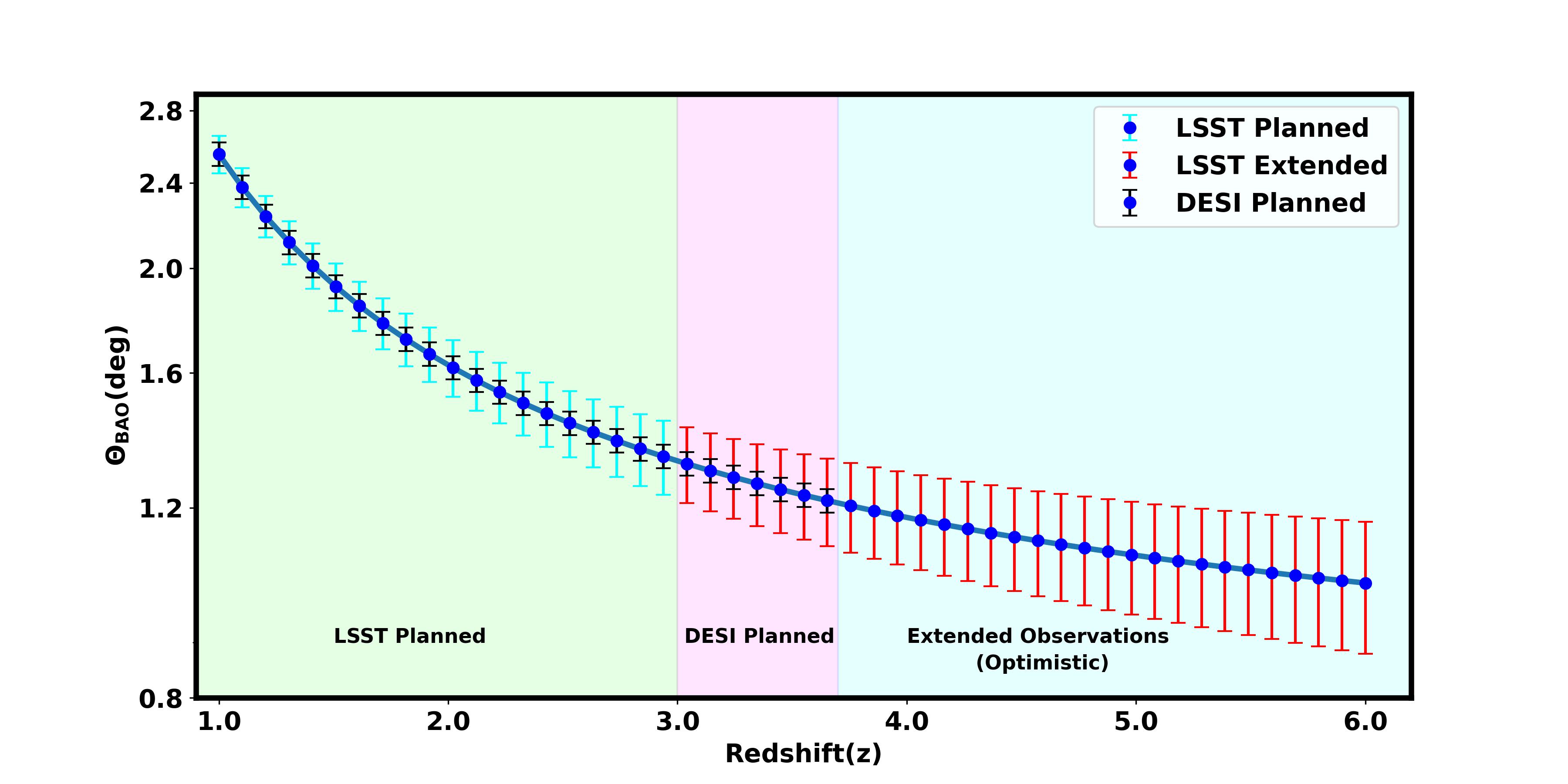}
\caption{This figure illustrates the evolution of the BAO scale and its associated error, primarily due to inaccuracies in photometric redshift (photo-z). For LSST's planned photo-z error, we show the error on BAO measurements from $\mathrm{z=1}$ to $\mathrm{z=3}$ in cyan. It also depicts the BAO measurement error expected from DESI up to redshift 3.7 (shown in black). Although there are no currently planned instruments to measure BAO beyond redshift 3.7, we extend the BAO measurements up to redshift 6 using the same photo-z error anticipated for Vera Rubin Observatory(in red). This visualization provides essential insights into the variability of the BAO scale and its error with increasing redshift, highlighting a rise in relative error as the standard deviation of the selection function increases with redshift.}
\label{fig:baoplot}
\end{figure}

Figure \ref{fig:baoplot} elaborates on the BAO scale's variation and associated error across different redshifts, emphasizing the influence of photo-z errors. For LSST's planned photo-z error, we illustrate the error on BAO measurements from $\mathrm{z=1}$ to $\mathrm{z=3}$ in cyan, showing how the expected errors will affect the precision of BAO measurements. Vera Rubin Observatory\cite{abell2009lsst, ivezic2019lsst}, along with projects like Euclid \cite{laureijs2011euclid}, will significantly enhance our understanding of BAO by surveying a total area of 18,000 square degrees, extending precise measurements to higher redshifts. The figure also depicts the BAO measurement error anticipated from Dark Energy Spectroscopic Instrument(DESI) \cite{collaboration2023early} up to redshift 3.7, shown in black, demonstrating the expected performance of DESI in this redshift range. The DESI is a notable five-year terrestrial survey focusing on exploring BAO and the evolution of cosmic structures through redshift-space distortions. Covering 14,000 square degrees of the sky, corresponding to a sky fraction ($\mathrm{f_{\text{sky}}}$) of 0.3, DESI aims for an accuracy better than 0.5\% within the redshift intervals $\mathrm{0.0 < z < 3.7}$.  Despite the lack of currently planned instruments to measure BAO beyond redshift 3.7, we extend the BAO measurements up to redshift 6 using the same photo-z error anticipated for LSST, shown in red. This extended range provides a hypothetical look at the potential capabilities of future surveys and their ability to probe deeper into the universe. This visualization provides essential insights into the variability of the BAO scale and its error with increasing redshift, highlighting a rise in relative error as the standard deviation of the selection function increases with redshift. With the capability of LISA in probing the high redshift Universe, this plot shows that future galaxy surveys with capabilities of measuring BAO beyond $\mathrm{z=3.5}$ can bring interesting tests of fundamental physics explored in this analysis.

\section{Forecast for reconstructing F(z) with redshift}
\label{sec:Forecast}

\subsection{Case with the fixed Hubble constant} 
\label{sec:fzfh}

The correlation between the EM luminosity distance at a specific redshift (z) and critical cosmological parameters, such as the BAO scale ($\mathrm{\theta_{BAO}}$) and the sound horizon ($\mathrm{r_s}$), is succinctly expressed by the Equation \ref{eq:f}. This equation is central to the analysis of the frictional term, linking it to three significant cosmological lengths: \texttt{(i)GW Luminosity Distance:} The distance for detectable GW events is estimated through the parameter estimation of GW events as described in Subsection \ref{sec:ParamEst}. \texttt{(ii)BAO Scale ($\mathrm{\theta_{BAO}}$):} Derived from the 2PACF and expressed in Equation \ref{eq:fit}, this scale is utilized in a versatile manner, allowing for application across various surveys. \texttt{(iii)Sound Horizon Distance:} Estimated by analyzing CMB acoustic oscillations through missions like WMAP  \cite{hinshaw2013nine}, Planck  \cite{aghanim2020planck}, ACTPol  \cite{thornton2016atacama}, SPT-3G  \cite{sobrin2022design}, and CMB-S4  \cite{abazajian2016cmb}. The first peak in the angular power spectrum at recombination reflects the sound horizon scale, measured precisely by Planck as $\mathrm{147.09 \pm 0.26}$ Mpc at the drag epoch and \texttt{redshift:} Inferred from the EM counterparts.

To estimate the posterior distribution of $\mathrm{\mathcal{F}(z)}$ as a function of redshift for $\mathrm{n_{GW}}$ GW sources, we employ a Hierarchical Bayesian framework. The posterior distribution is represented as follows \cite{Afroz:2023ndy}

\begin{equation}
    \mathrm{P(\mathcal{F}(z)) \propto  \Pi(\mathcal{F}(z))\prod_{i=1}^{n_{GW}}\iiint dr_sP(r_s) dD_L^{GW^i}d\theta_{BAO}^i P(\theta_{BAO}^i) \mathcal{L}(D_L^{GW^i}|\mathcal{F}(z^i),\theta_{BAO}^i,r_s,z^i)}.
\label{eq:FZposterior}
\end{equation}

In this framework, $\mathrm{P(\mathcal{F}(z))}$ denotes the posterior probability density of $\mathrm{\mathcal{F}(z)}$. The term $\mathrm{\mathcal{L}(D_L^{\text{GW}^i}|\mathcal{F}(z^i), \theta_{\text{BAO}}^i, r_s, z^i)}$ corresponds to the likelihood function, which represents the posterior distribution of the GW luminosity distance. In this expression, $\mathrm{P(r_s)}$  and $\mathrm{P(\theta_{BAO}^i)}$ representing the prior probabilities for the sound horizon and BAO scale, respectively. $\mathrm{\Pi(\mathcal{F}(z))}$ indicates the prior distribution for $\mathrm{\mathcal{F}(z)}$, reflecting pre-existing knowledge or assumptions about the frictional term before the data is considered. A flat prior from 0.1 to 2 is adopted for $\mathrm{\mathcal{F}(z)}$. This investigation focuses solely on bright sirens, assuming that the redshift of the host galaxies of the GW sources is measured spectroscopically. This assumption implies precise measurements of redshift for these sources, as the error in redshift measurements is very small compared to the errors associated with the BAO scale and distance measurements.

\subsection{Case with varying the Hubble constant}
\label{sec:fzvh0}
In previous discussions, we concentrated on analyzing the frictional term, denoted as $\mathrm{\mathcal{F}(z)}$ in Equation \ref{eq:f}, under the assumption of a constant Hubble constant ($\mathrm{H_0}$). Moving forward, we expand our analysis to concurrently estimate both $\mathrm{\mathcal{F}(z)}$ and $\mathrm{H_0}$. Our primary objective remains the precise determination of $\mathrm{\mathcal{F}(z)}$, while concurrently assessing $\mathrm{H_0}$. This dual estimation strategy is especially pertinent with the advent of future detectors, which will enhance our capability to probe deeper into redshifts, thus enabling either combined or independent measurements of these parameters. This methodology is particularly promising for addressing the so-called Hubble tension (for a comprehensive review \cite{Abdalla:2022yfr}), the notable discrepancy between the locally measured values of $\mathrm{H_0}$ and those inferred from CMB observations \cite{Planck:2018vyg}. 

By employing a hierarchical Bayesian framework, we efficiently estimate these parameters using a comprehensive array of observational data, encompassing GW detections, CMB data, and large-scale structure surveys, as elaborated in earlier sections. The hierarchical integration of these datasets is meticulously designed to manage their distinct uncertainties and biases effectively. This integration produces correlated constraints that significantly enhance our understanding of cosmological dynamics, help resolve discrepancies among observational datasets, and uncover critical relationships between key cosmological parameters. The joint posterior distribution for $\mathrm{\mathcal{F}(z)}$ and $\mathrm{H_0}$, based on these considerations, is formulated as follows
\begin{equation}
\begin{aligned}
    &\mathrm{P(\mathcal{F}(z),H_0) \propto  \Pi(\mathcal{F}(z))\Pi(H_0)\prod_{i=1}^{n_{GW}}\iiint}  \mathrm{dr_sP(r_s^i) dD_L^{GW^i} 
    d\theta_{BAO}^i  P(\theta_{BAO}^i) \mathcal{L}(D_L^{GW^i}|\mathcal{F}(z^i),H_0,\theta_{BAO}^i,r_s,z^i)}.
\end{aligned}
\label{FZ}
\end{equation}

In this model, $\mathrm{P(\mathcal{F}(z), H_0})$ denotes the joint posterior probability density function for $\mathrm{\mathcal{F}(z)}$ and $\mathrm{H_0}$. The likelihood function is represented by $\mathrm{\mathcal{L}(D_L^{\text{GW}^i}|\mathcal{F}(z)^i, H_0^i, \theta_{\text{BAO}^i}, r_s, z^i)}$. The terms $\mathrm{\Pi(\mathcal{F}(z))}$ and $\mathrm{\Pi(H_0)}$ define the prior distributions for $\mathrm{\mathcal{F}(z)}$ and $\mathrm{H_0}$, respectively, encapsulating pre-existing beliefs or assumptions prior to analyzing the data. Within our hierarchical Bayesian framework, we utilize flat priors for both parameters, indicating minimal initial bias. The range for $\mathrm{\mathcal{F}(z)}$ is set from 0.1 to 2.0, and for $\mathrm{H_0}$, from 40 km/s/Mpc to 100 km/s/Mpc.

\section{Results}
\label{sec:result}

\begin{figure}[ht]
\centering
\includegraphics[height=6.5cm, width=16cm]{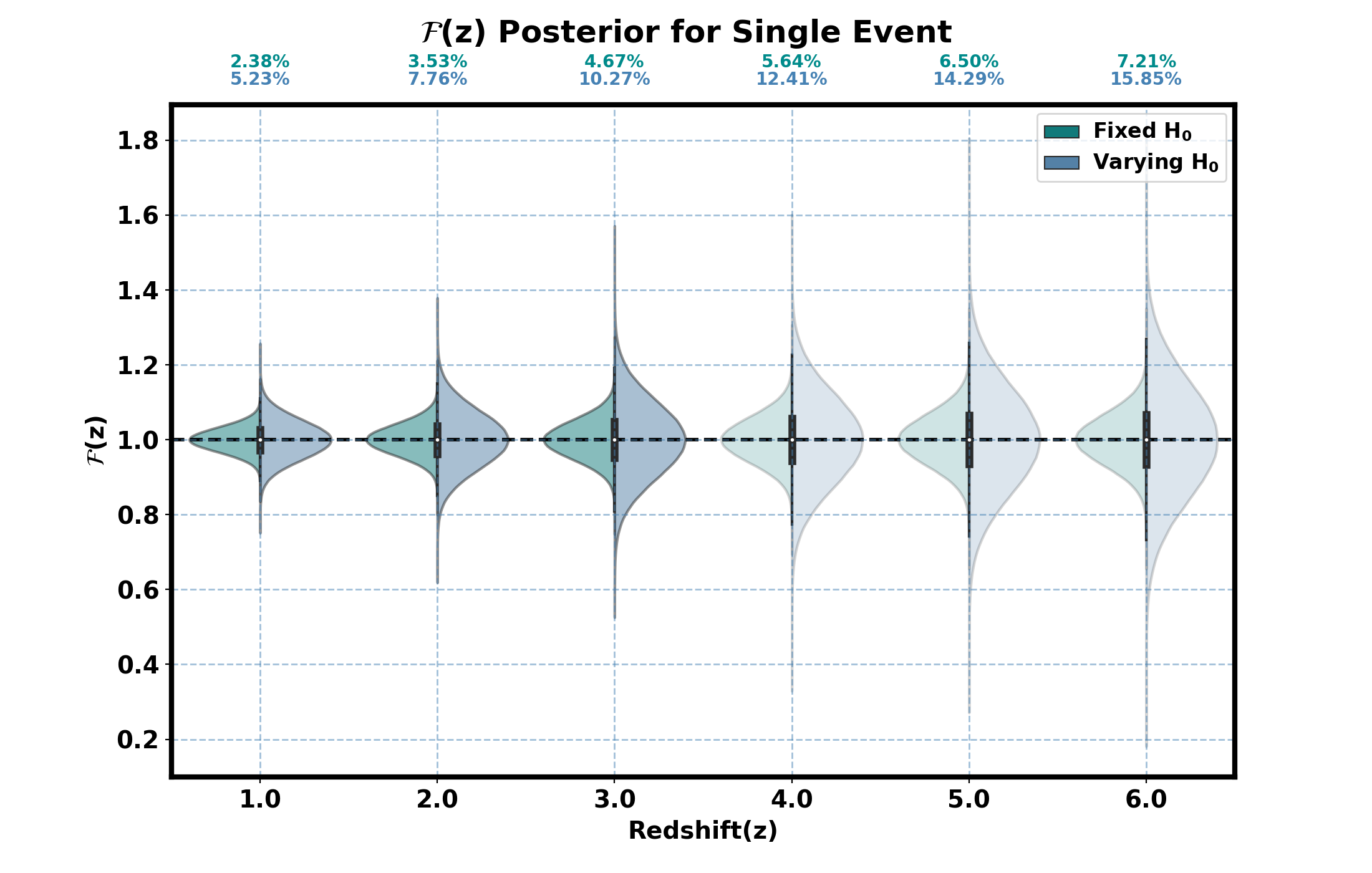}
\caption{This violin plot depicts the posterior distribution of the non-GR parameter $\mathrm{\mathcal{F}(z)}$, illustrating its variation with cosmic redshift for LISA across both constant and varying $\mathrm{H_0}$ scenarios for a single event. The upper x-axis shows the corresponding 1-$\mathrm{\sigma}$ error bars for each measurement. Beyond redshift $\mathrm{z=3.0}$, the plot uses a fainter color to indicate that the BAO scale cannot be measured beyond this redshift with currently planned galaxy surveys. A line at $\mathrm{\mathcal{F}(z)=1}$ is included to represent the GR case. This figure underscores the enhanced capability of the future space-based detector LISA to measure $\mathrm{\mathcal{F}(z)}$ with increased precision across extended cosmic distances. The plot is based on a single MBBH merger event with a fixed total mass of $\mathrm{10^6}$ $\mathrm{M_{\odot}}$ having a mass ratio $q=1$.}
\label{fig:violin}
\vspace{-0.1cm}
\end{figure}

\begin{figure}[ht]
    \centering
    \includegraphics[height=7.0cm, width=16cm]{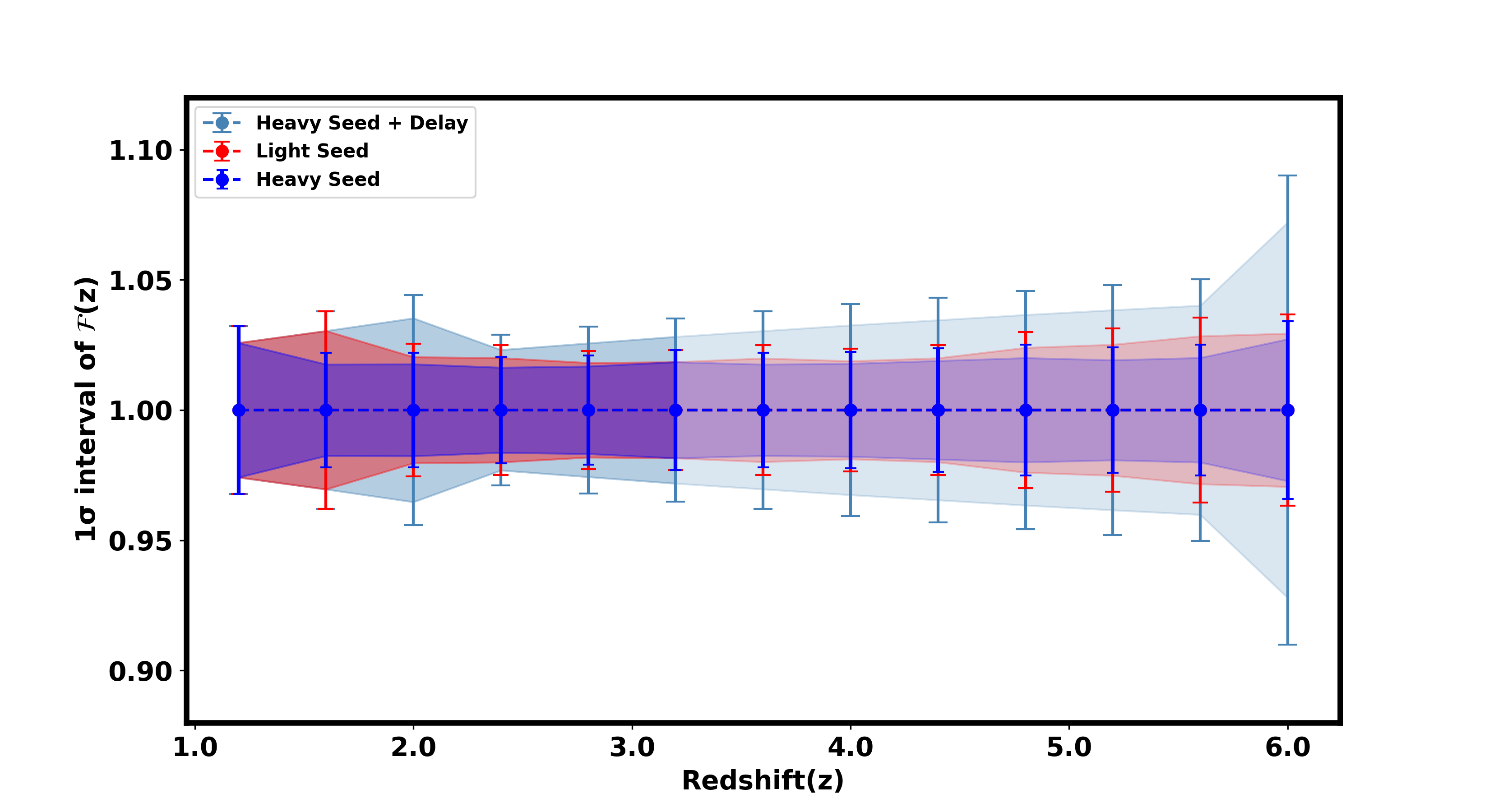}
    \caption{This plot illustrates the precision in reconstructing the redshift evolution of the Planck mass, represented by $\mathcal{F}(z)$, for the Light Seed, Heavy Seed, and Heavy Seed+delay models using the upcoming space-based LISA observatory. The observation time is 4 years for the Light Seed and Heavy Seed models and 10 years for the Heavy Seed+delay model. The filled regions represent the $1\sigma$ intervals for the assumption of $75\%$ EM counterpart detection, while the error bars with lines indicate the $1\sigma$ intervals for the assumption of $50\%$ EM counterpart detection case. The plot highlights LISA's accuracy in measuring the function $\mathcal{F}(z)$ up to a redshift of 6. Beyond a redshift of 3.0, a fainter color is used to indicate that the BAO scale cannot be measured beyond this redshift with the currently planned galaxy surveys.}
    \label{fig:Fzreconstruction}
\end{figure}

This study exclusively examines bright sirens detectable by the space-based GWs observatory LISA, focusing on events with a redshift up to $\mathrm{z=6}$ across multiple population models. We present the frictional term, $\mathrm{\mathcal{F}(z)}$, measurements under two scenarios: with a fixed Hubble constant ($\mathrm{H_0}$) and with a varying $\mathrm{H_0}$. The comparative results are illustrated in Figure \ref{fig:violin} and Figure \ref{fig:Fzreconstruction}. Specifically, Figure \ref{fig:violin} showcases the impact of single-event measurements, and Figure \ref{fig:Fzreconstruction} highlights how event detection rates and population assumptions influence the precision of $\mathrm{\mathcal{F}(z)}$ measurements.

\subsection{ Reconstruction of F(z)}

In cases where  $\mathrm{H_0}$ is held constant, it acts as an overall normalization factor for $\mathrm{\mathcal{F}(z)}$ across different redshifts. This setup provides a unique opportunity to normalize $\mathrm{\mathcal{F}(z)}$ measurements to 1 if an independent and precise measurement of $\mathcal{F}(z)$ in our local universe is possible. Such normalization, particularly effective at low redshifts, helps eliminate the dependence on $\mathrm{H_0}$ and demonstrates the potential deviations from GR due to the cumulative effects of modified gravity wave propagation over cosmological distances. The uncertainty in the measurement of the frictional term, $\mathrm{\mathcal{F}(z)}$, is inversely proportional to the square root of the number of GW sources, denoted as $\mathrm{1/\sqrt{N_{GW}}}$. This relationship suggests that increasing the count of GW sources detected by LISA enhances the precision of $\mathrm{\mathcal{F}(z)}$ measurements. However, the accuracy improvement reaches a plateau beyond a certain number of sources, primarily due to errors associated with the BAO scale. The precision in estimating the BAO scale emerges as a critical factor limiting the refinement of $\mathrm{\mathcal{F}(z)}$ estimation accuracy. Thus, advancing the accuracy of BAO scale measurements is essential for further enhancements in determining the frictional term with redshift. Essentially, optimizing the number of GW observations, alongside improvements in BAO scale accuracy and GWs luminosity distance measurements, is crucial for advancing our understanding of $\mathrm{\mathcal{F}(z)}$ and probing new aspects of fundamental physics.

On the GWs front, the accuracy of $\mathrm{\mathcal{F}(z)}$ relies predominantly on the precision of the luminosity distance to GWs sources, $\mathrm{D_L^{GW}(z)}$. Enhancements in measuring $\mathrm{D_L^{GW}(z)}$ would significantly boost the fidelity of $\mathrm{\mathcal{F}(z)}$ estimations. Given the degeneracy of $\mathrm{D_L^{GW}(z)}$ with other GWs parameters, notably the inclination angle ($\mathrm{i}$), acquiring more accurate measurements of the inclination angle, potentially through EM counterparts, could substantially reduce uncertainties in $\mathrm{D_L^{GW}(z)}$ and, by extension, refine $\mathrm{\mathcal{F}(z)}$ estimates  \cite{xie2023breaking}. The effect of peculiar velocities on redshift inference from EM counterparts is notably critical at lower redshifts ($\mathrm{z<0.03}$) but becomes negligible at higher redshifts  \cite{Mukherjee:2019qmm}. Given that the majority of GW sources detected by LISA are expected to be at higher redshifts, this effect is disregarded in our analysis.

\subsection{ Reconstruction of F(z) and Hubble Constant}

Building upon the hierarchical Bayesian framework introduced earlier to estimate $\mathrm{\mathcal{F}(z)}$ with a fixed $\mathrm{H_0}$, we now extend this approach to concurrently measure $\mathrm{\mathcal{F}(z)}$ and the Hubble constant, $\mathrm{H_0}$. In this extended framework, we derive the EM luminosity distance by integrating various cosmological scales, including the BAO scale, sound horizon distance, and redshift. Figures \ref{fig:violin} provide a detailed analysis of $\mathrm{\mathcal{F}(z)}$ measurements under scenarios of both fixed and varying $\mathrm{H_0}$ with only six events up to $z=6$. The uncertainty on the parameter $\mathrm{\mathcal{F}(z)}$ increased by approximately a factor of 2 to 2.5 when $\mathrm{H_0}$ is allowed to vary relative to scenarios with a fixed $\mathrm{H_0}$. This scenario with only six events corresponds to a pessimistic case with only $4\%$ EM bright events for the Light Seed model and the Heavy Seed model and about $30\%$ for the Heavy Seed+Delay model in four years of observation period. This corresponds to a single events at every redshift up to $z=6$. We have shown the constraints for a few scenarios of detectable bright sirens from LISA in appendix \ref{sec:ErrorBudget}. If the efficiency of LISA bright sirens are less than this, then one needs to consider a longer observation period of the LISA mission for at least one detection of bright standard siren.

\begin{figure}[ht]
    \centering
    \includegraphics[height=8.5cm, width=16cm]{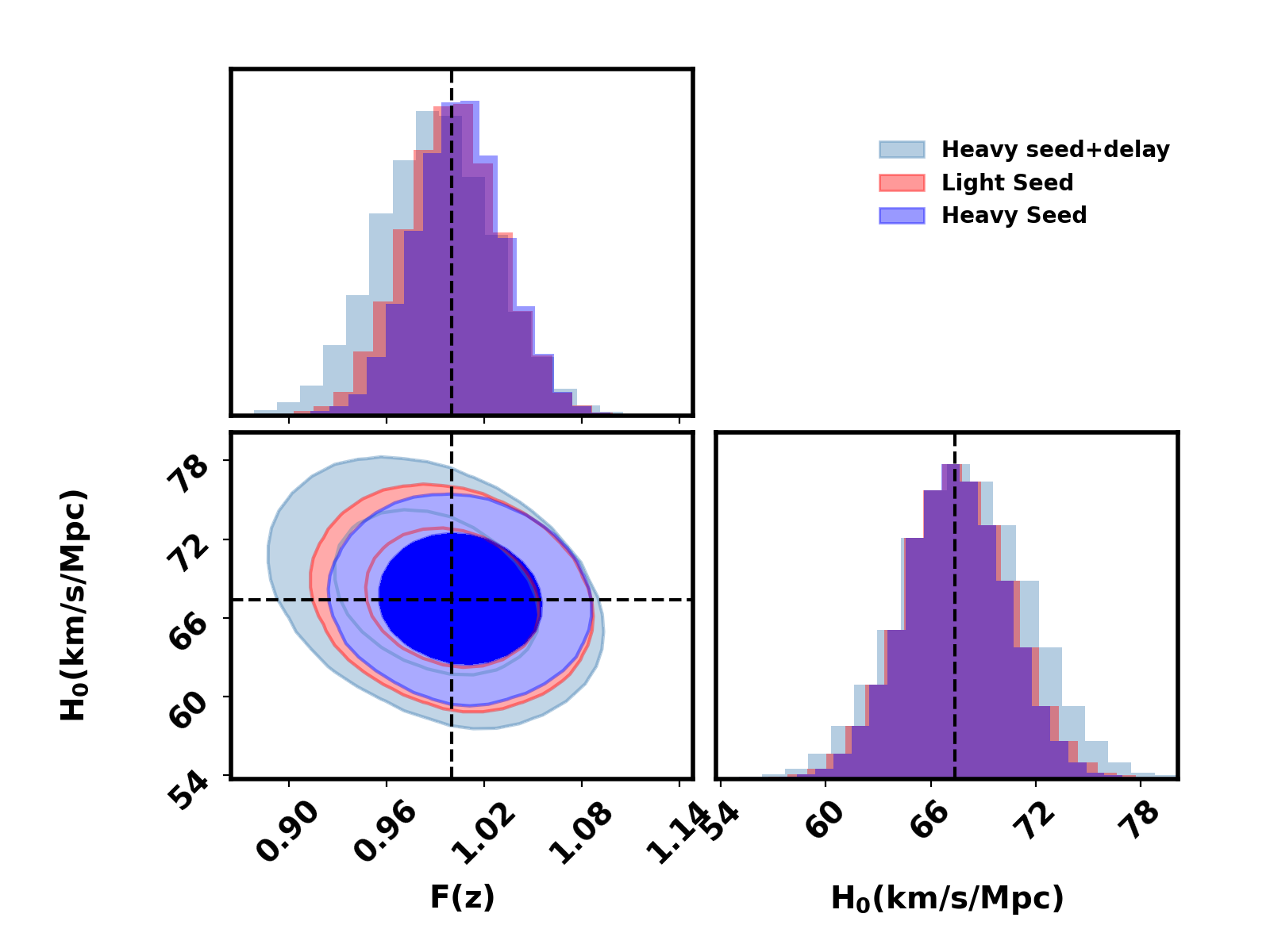}
    \caption{This plot illustrates the optimal combined measurements of $\mathrm{\mathcal{F}(z)}$ and the Hubble constant ($\mathrm{H_0}$) for the Light Seed, Heavy Seed, and Heavy Seed+delay models using the forthcoming space-based LISA observatory. The observation periods are 4 years for the Light Seed and Heavy Seed models, and 10 years for the Heavy Seed+delay model, assuming that EM counterparts are detected for $\mathrm{75\%}$ of the events. These optimal measurements occur at redshifts 3.2, 4.0, and 2.4 for the Light Seed, Heavy Seed, and Heavy Seed+delay models, respectively.}
    \label{fig:CombinedPost}
\end{figure}

\begin{figure}
    \centering
    \includegraphics[height=7.0cm, width=16cm]{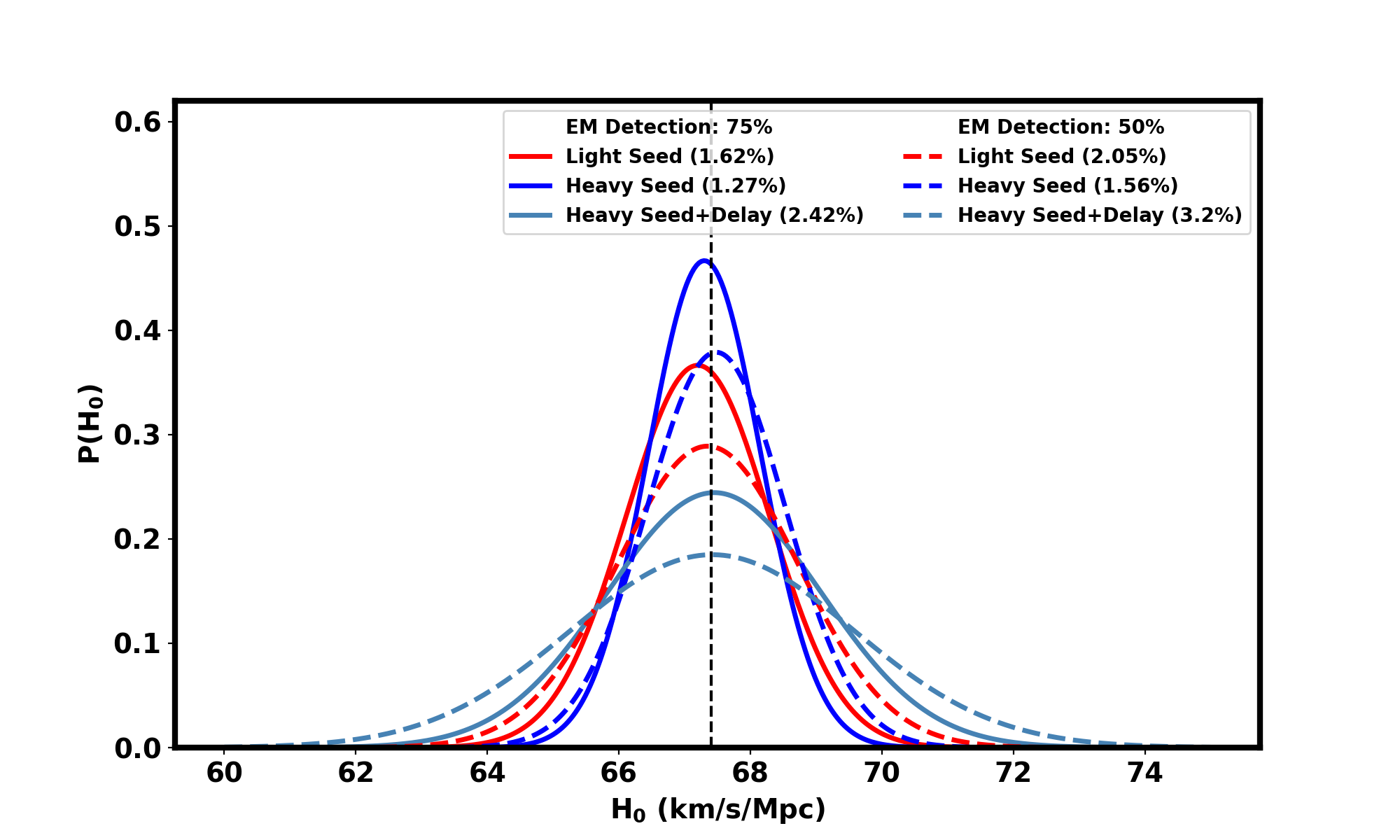}
    \caption{This plot shows the posterior distribution of the Hubble constant ($\mathrm{H_0}$) marginalized over $\mathrm{\mathcal{F}(z)}$ for all events up to a redshift of 6, based on three models: Light Seed, Heavy Seed, and Heavy Seed+delay, using the upcoming space-based LISA observatory. The observation periods are 4 years for both the Light Seed and Heavy Seed models and 10 years for the Heavy Seed+delay model. The plot includes results assuming that EM counterparts are detected for 75\% of the events (solid lines) and 50\% of the events (dashed lines). The 1$\mathrm{\sigma}$ uncertainties in the measurements are provided in parentheses in the legend.}
    \label{fig:H0posterior}
\end{figure}

In Figure \ref{fig:Fzreconstruction}, we specifically illustrate the improved precision in tracking the redshift-dependent variations of the Planck mass ($\mathrm{\mathcal{F}(z)}$) for MBBHs systems, emphasizing enhanced measurements in the redshift range from $z=1.2$ to $z=6.0$. Additionally, Figure \ref{fig:CombinedPost} illustrates the optimal joint measurements of $\mathrm{\mathcal{F}(z)}$ and the Hubble constant ($\mathrm{H_0}$). In Figure \ref{fig:H0posterior}, we show the measurement of $H_0$ marginalized over $\mathrm{\mathcal{F}(z)}$ for the three different population models, for all events up to a redshift of $\mathrm{z=6.0}$. The results on $\mathrm{\mathcal{F}(z)}$ and $H_0$ for scenario with MBBHs distribution having a mass ratio beyond $q=1$ is shown in the appendix \ref{sec:appendix}.

\section{Conclusion \& Discussion}
\label{sec:conclusion}

In this study, we introduce a model-independent approach for investigating the frictional term in theories of gravity, capable of identifying deviations from GR in the cosmological scales across any redshift. This analysis is facilitated by bright sirens that will be observable by the upcoming space-based GWs detector LISA, which is designed to operate in the milliHertz frequency range. Building on our previous work  \cite{Afroz:2023ndy,Afroz:2024joi}, we have already demonstrated a model-independent reconstruction of the frictional term using current and forthcoming ground-based GW detectors, which primarily operate in the hertz to kilohertz frequency spectrum.

We propose a novel strategy that integrates three pivotal cosmological measures as functions of redshift: the luminosity distance from GW sources $\mathrm{D_L^{\mathrm{GW}}(z)}$, the angular scale of the BAO from galaxy surveys $\mathrm{\theta_{BAO}(z)}$, and the sound horizon $\mathrm{r_s}$ from CMB observations. This amalgamation provides a unique framework to scrutinize alternate theories of gravity, particularly focusing on the frictional term. Through an integrated analysis combining BAO measurements, sound horizon distances, redshiftd from EM counterpart, and GW luminosity distances, we enable a robust measurement of cosmological parameters. This approach notably improves our ability to simultaneously determine the Hubble constant ($\mathrm{H_0}$) and the frictional term as a function of redshift in a model-independent way. We have demonstrated the feasibility of achieving precise measurements of the frictional term with an accuracy ranging from $\mathrm{2.38\%}$ to $\mathrm{7.21\%}$ up to a redshift of 6, utilizing just a single GW source detected by LISA. Additionally, we achieve a measurement of the Hubble constant with a precision of $\mathrm{1.27\%}$, marginalized over the measurements of deviations in the frictional term, for 4 years of observation time. This assumes that EM are detected for $\mathrm{75\%}$ of the events. The accuracy of these measurements scales inversely with the square root of the number of events, which is proportional to the square root of the observation time, expressed as $\mathrm{1/\sqrt{T_{\mathrm{obs}}/\text{years}}}$.  Furthermore, our methodology allows for the reconstruction of the frictional term as a function of redshift without relying on any specific parametric forms, highlighting the strength and flexibility of this model-independent approach in advancing our understanding of cosmic expansion and the underlying forces.

The current BAO measurements from the Vera Rubin Observatory, Euclid, and DESI, which cover redshifts up to z=3.7. This limitation on the BAO measurements impacts our results on the reconstruction of the frictional term as a function of cosmic redshift. This limitation is illustrated in Figures \ref{fig:violin} and \ref{fig:Fzreconstruction}, where we show the results using deeper colors for redshifts up to $\mathrm{z=3.0}$. Beyond redshift $\mathrm{z=3.0}$, the results are shown in fainter colors to highlight the limitations of our current planned BAO measurements, as we have used LSST-like measurements for our analysis. Improved surveys with deeper reach could potentially overcome these limitations, providing more accurate reconstructions and deeper insights into the frictional term across a wider range of redshifts.

Additionally, our proposed methodology can be adapted to incorporate dark sirens, where accurately determining the redshift is critical. For this, the three-dimensional clustering of GW sources with galaxies is employed to deduce redshifts, as highlighted in prior studies.  \cite{mukherjee2018beyond, mukherjee2020probing, Mukherjee:2020hyn, Mukherjee:2022afz, abbott2021constraints,Afroz:2024joi}. It is important to note that the current analysis focuses strictly on bright sirens. This focus necessitates the exclusion of certain GW sources such as EMRIs and stellar mass BHBs, which typically do not have EM counterparts. Furthermore, our study limits its scope to redshifts up to $\mathrm{z=6}$ due to the increasing difficulty in detecting EM counterparts beyond this point. Incorporating dark sirens, which do not have EM counterparts, could significantly refine the accuracy of our frictional term measurements. The extension of our approach to include dark sirens could lead to more precise determinations of the frictional term, thereby enhancing our understanding of the fundamental principles of gravity.

\section*{Acknowledgements}

This work is part of the \texttt{⟨data|theory⟩ Universe-Lab}, supported by TIFR and the Department of Atomic Energy, Government of India. We would like to thank Enrico Barausse for providing comments on the draft. We also express gratitude to the computer cluster of \texttt{⟨data|theory⟩ Universe-Lab} and the TIFR computer center HPC facility for computing resources. We acknowledge the use of the following packages in this work: Astropy \cite{robitaille2013astropy, price2018astropy}, Bilby \cite{ashton2019bilby},
Pandas \cite{mckinney2011pandas}, NumPy \cite{harris2020array}, Seaborn \cite{bisong2019matplotlib}, CAMB \cite{lewis2011camb}, Scipy \cite{virtanen2020scipy}, Dynesty \cite{speagle2020dynesty}, emcee \cite{foreman2013emcee}, Matplotlib \cite{Hunter:2007} and BBHx \cite{Katz:2020hku}.

\appendix

\section{Impact of Electromagnetic Counterpart Detection on $\mathrm{\mathcal{F}(z)}$ and $\mathrm{H_0}$ Measurements}
\label{sec:ErrorBudget}
In this section, we present a few cases of bright sirens detectable from LISA and summarize the best measurements of $\mathrm{\mathcal{F}(z)}$ and the combined error on $\mathrm{H_0}$ for three different models: Light Seed, Heavy Seed, and Heavy Seed + Delay. Each table provides results for various event selections, including 75\%, 50\%, 25\%, and 10\% of selected events. The tables detail the maximum redshift up to which a source is detected in LISA ($\mathrm{z_{max}}$), the total number of GW events ($\mathrm{N_{GW}}$) up to $\mathrm{z_{max}}$, and the associated precision on $\mathrm{\mathcal{F}(z)}$ and $H_0$. For $\mathrm{\mathcal{F}(z)}$ we quote the minimum uncertainty and corresponding redshift, and for the Hubble constant, we mention the uncertainty after combining $\mathrm{N_{GW}}$ events.  

Tables \ref{tab:LSerror},  \ref{tab:HSerror}, and  \ref{tab:HSDerror} present the results for the Light Seed model,  Heavy Seed model and the Heavy Seed + Delay model respectively. For each model, the tables highlight the least percentage error in $\mathrm{\mathcal{F}(z)}$ at specific redshifts, and the overall combined error on $\mathrm{H_0}$. Note that for the 10\% selection case, where no events are available, it is denoted as "--". These tables illustrate how the precision of $\mathrm{\mathcal{F}(z)}$ and $\mathrm{H_0}$ varies with the number of selected events and their redshift range, reflecting the impact of different selection thresholds on measurement accuracy.

\begin{table}
\centering
\begin{tabular}{|l|c|c|c|c|}
\hline
Bright events & $\mathrm{z_{max}}$ & $\mathrm{N_{GW}}$ & Least uncertainty on $\mathrm{\mathcal{F}(z)}$ & Combined uncertainty on $\mathrm{H_0}$   \\
\hline
75\% Selected & 6 & 73 & 2.65\% at z=1.2 & 1.62\% \\
\hline
50\% Selected & 6 & 49 & 2.65\% at z=1.2 & 2.05\%  \\
\hline
25\% Selected & 6 & 23 & 3.57\% at z=2.0 & 3.40\%  \\
\hline
10\% Selected & 5.2 & 6 & 5.02\% at z=3.2 & 6.42\%  \\
\hline
\end{tabular}
\caption{Table showing the best measurement of $\mathrm{\mathcal{F}(z)}$ and the combined error on $\mathrm{H_0}$ for the Light Seed model. The table summarizes results for various event selections, detailing the maximum redshift ($\mathrm{z_{max}}$), the number of GW events ($\mathrm{N_{GW}}$), and the associated precision for 75\%, 50\%, 25\%, and 10\% of selected events. For each selection, the table highlights the least percentage error in $\mathrm{\mathcal{F}(z)}$ at a specific redshift and the overall combined error on $\mathrm{H_0}$. These results illustrate how the precision of $\mathrm{\mathcal{F}(z)}$ and $\mathrm{H_0}$ varies with the number of selected events and their redshift range, reflecting the impact of different selection thresholds on measurement accuracy.}

\label{tab:LSerror}
\end{table}

\begin{table}
\centering
\begin{tabular}{|l|c|c|c|c|}
\hline
Bright events & $\mathrm{z_{max}}$ & $\mathrm{N_{GW}}$ & Least uncertainty on $\mathrm{\mathcal{F}(z)}$ & Combined uncertainty on $\mathrm{H_0}$   \\
\hline
75\% Selected & 6 & 98 & 2.62\% at z=1.2 & 1.27\% \\
\hline
50\% Selected & 6 & 66 & 2.62\% at z=1.2 & 1.56\%  \\
\hline
25\% Selected & 6 & 32 & 3.02\% at z=1.6 & 2.60\%  \\
\hline
10\% Selected & 6 & 9 & 4.51\% at z=2.8 & 4.50\%  \\
\hline
\end{tabular}
\caption{Table showing the best measurement of $\mathrm{\mathcal{F}(z)}$ and the combined error on $\mathrm{H_0}$ for the Heavy Seed model. The table summarizes results for various event selections, detailing the maximum redshift ($\mathrm{z_{max}}$), the number of GW events ($\mathrm{N_{GW}}$), and the associated precision for 75\%, 50\%, 25\%, and 10\% of selected events. For each selection, the table highlights the least percentage error in $\mathrm{\mathcal{F}(z)}$ at a specific redshift and the overall combined error on $\mathrm{H_0}$. These results illustrate how the precision of $\mathrm{\mathcal{F}(z)}$ and $\mathrm{H_0}$ varies with the number of selected events and their redshift range, reflecting the impact of different selection thresholds on measurement accuracy.}

\label{tab:HSerror}
\end{table}

\begin{table}
\centering
\begin{tabular}{|l|c|c|c|c|}
\hline
Bright events & $\mathrm{z_{max}}$ & $\mathrm{N_{GW}}$ & Least uncertainty on $\mathrm{\mathcal{F}(z)}$ & Combined uncertainty on $\mathrm{H_0}$   \\
\hline
75\% Selected & 6 & 31 & 2.63\% at z=1.2 & 2.42\% \\
\hline
50\% Selected & 6 & 21 & 2.63\% at z=1.2 & 3.20\%  \\
\hline
25\% Selected & 5.6 & 9 & 4.26\% at z=2.4 & 6.82\%  \\
\hline
10\% Selected & -- & -- & -- & --  \\
\hline
\end{tabular}
\caption{Table showing the best measurement of $\mathrm{\mathcal{F}(z)}$ and the combined error on $\mathrm{H_0}$ for the Heavy Seed + Delay model. The table summarizes results for various event selections, detailing the maximum redshift ($\mathrm{z_{max}}$), the number of GW events ($\mathrm{N_{GW}}$), and the associated precision for 75\%, 50\%, 25\%, and 10\% of selected events. For each selection, the table highlights the least percentage error in $\mathrm{\mathcal{F}(z)}$ at a specific redshift and the overall combined error on $\mathrm{H_0}$. Note that for the 10\% case, there are no events, so it is denoted as "--". These results illustrate how the precision of $\mathrm{\mathcal{F}(z)}$ and $\mathrm{H_0}$ varies with the number of selected events and their redshift range, reflecting the impact of different selection thresholds on measurement accuracy.}
\label{tab:HSDerror}
\end{table}

\section{Measurements for Different Mass Ratio Distributions}
\label{sec:appendix}

In this section, we extend our analysis of $\mathcal{F}(z)$ and $H_0$ using LISA bright sirens by considering different mass ratio distributions for binary mergers. Previously, we presented results assuming mass ratio $q = 1$. Here, we explore the impact of varying mass ratios on the measurements of $\mathcal{F}(z)$ and $H_0$. Specifically, we investigate mass ratio distributions following $\mathrm{q^2}$, where $\mathrm{q}$ is defined as the ratio of the component masses and ranges from $\mathrm{10^{-3}}$ to 1. We consider two selection criteria for these distributions: 75\% and 50\%, corresponding to the different proportions of systems retained based on their mass ratios. This approach allows us to understand how variations in the mass ratio distribution affect the detectability and measurement precision of bright sirens.

In Table \ref{tab:EventSummaryq2} we provides a summary of the total number of events and the number of detected events for different merger models with an SNR threshold of 50, categorized by the mass ratio selection criteria of 75\% and 50\%. The results show that the distribution of mass ratios does not significantly affect the number of detected events, as our study is limited to sufficiently low redshifts where most events are very bright. Comparing these findings with the results in Table \ref{tab:EventSummary}, we find that the number of detected events differs by approximately 2-7\%, depending on the model. This variation is most pronounced in the high-redshift portion of our analysis.

\begin{table}
\centering
\begin{tabular}{|l|c|c|c|c|c|}
\hline
Model & $T_{obs}$ (years) & Total Events & Selected (75\%) & Selected (50\%) & 1 Event Detectability  \\
\hline
LS & 4 & 158 & 68 & 44 & 6.82\% \\
\hline
HS & 4 & 147 & 96 & 64 & 4.69\% \\
\hline
HSD & 10 & 49 & 29 & 20 & 15.01\% \\
\hline
\end{tabular}
\caption{Number of total and detected events for different merger models Light Seed (LS), Heavy Seed (HS), and Heavy Seed+Delay (HSD), with a threshold SNR of 50. The table includes results based on 75\% and 50\% selection criteria, as well as the percentage of detectability of single events at each integer redshift from 1 to 6 (1 Event Detectability).}
\label{tab:EventSummaryq2}
\end{table}

\begin{figure}[ht]
    \centering
    \includegraphics[height=7.0cm, width=16cm]{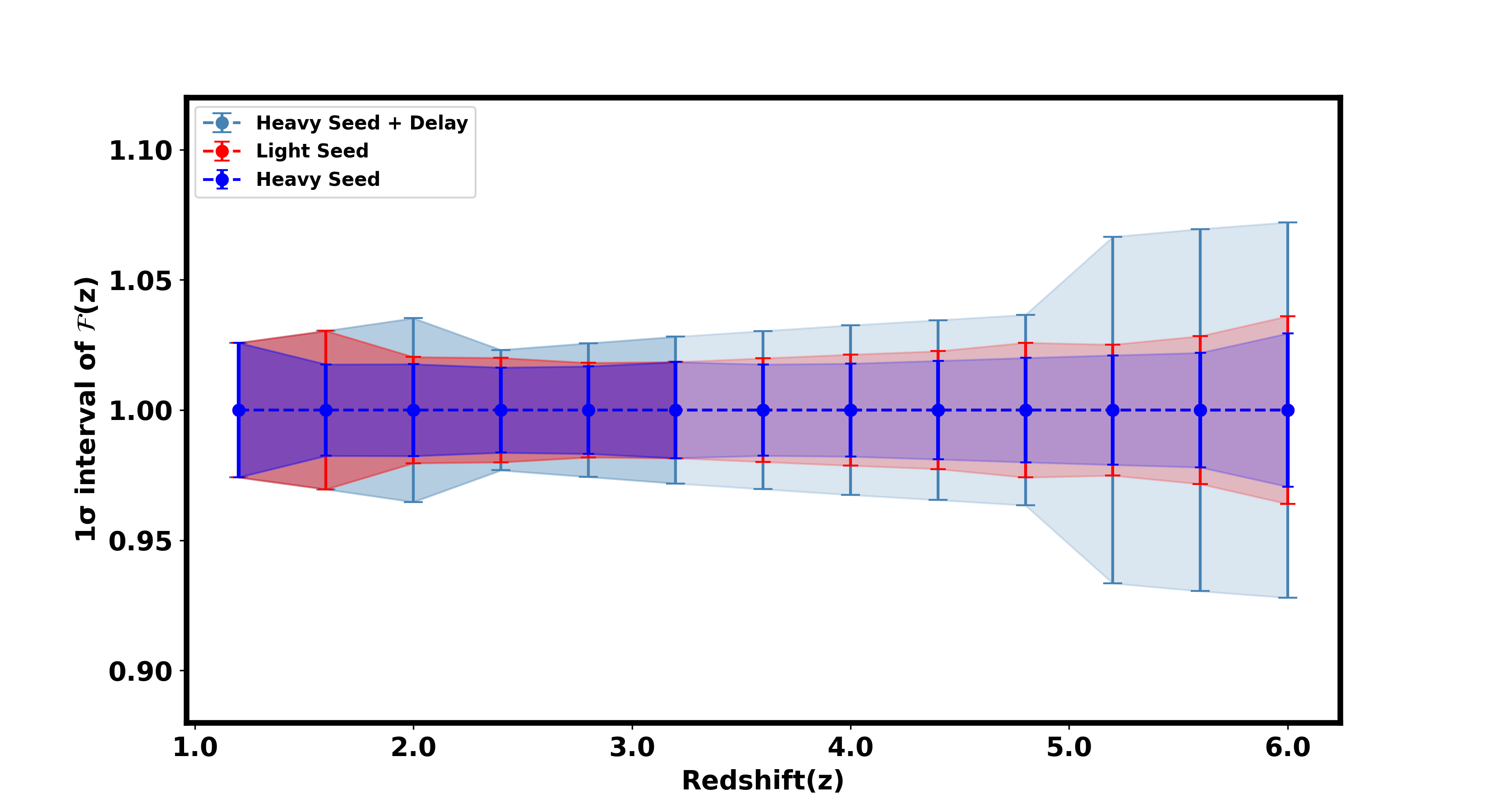}
    \caption{This plot illustrates the precision in reconstructing the redshift evolution of the Planck mass, represented by $\mathrm{\mathcal{F}(z)}$, for the Light Seed, Heavy Seed, and Heavy Seed+delay models using the upcoming space-based LISA observatory. The observation time is 4 years for the Light Seed and Heavy Seed models and 10 years for the Heavy Seed+delay model, assuming the $\mathrm{q^2}$ mass ratio distribution and that EM counterparts are detected for 75\% of the events. The plot highlights LISA's accuracy in measuring $\mathrm{\mathcal{F}(z)}$ up to a redshift of 6. Beyond a redshift of 3.0, a fainter color is used to indicate that the BAO scale cannot be measured beyond this redshift with the currently planned galaxy surveys.}
    \label{fig:Fzreconstructionq2}
\end{figure}

\begin{figure}
    \centering
    \includegraphics[height=7.0cm, width=16cm]{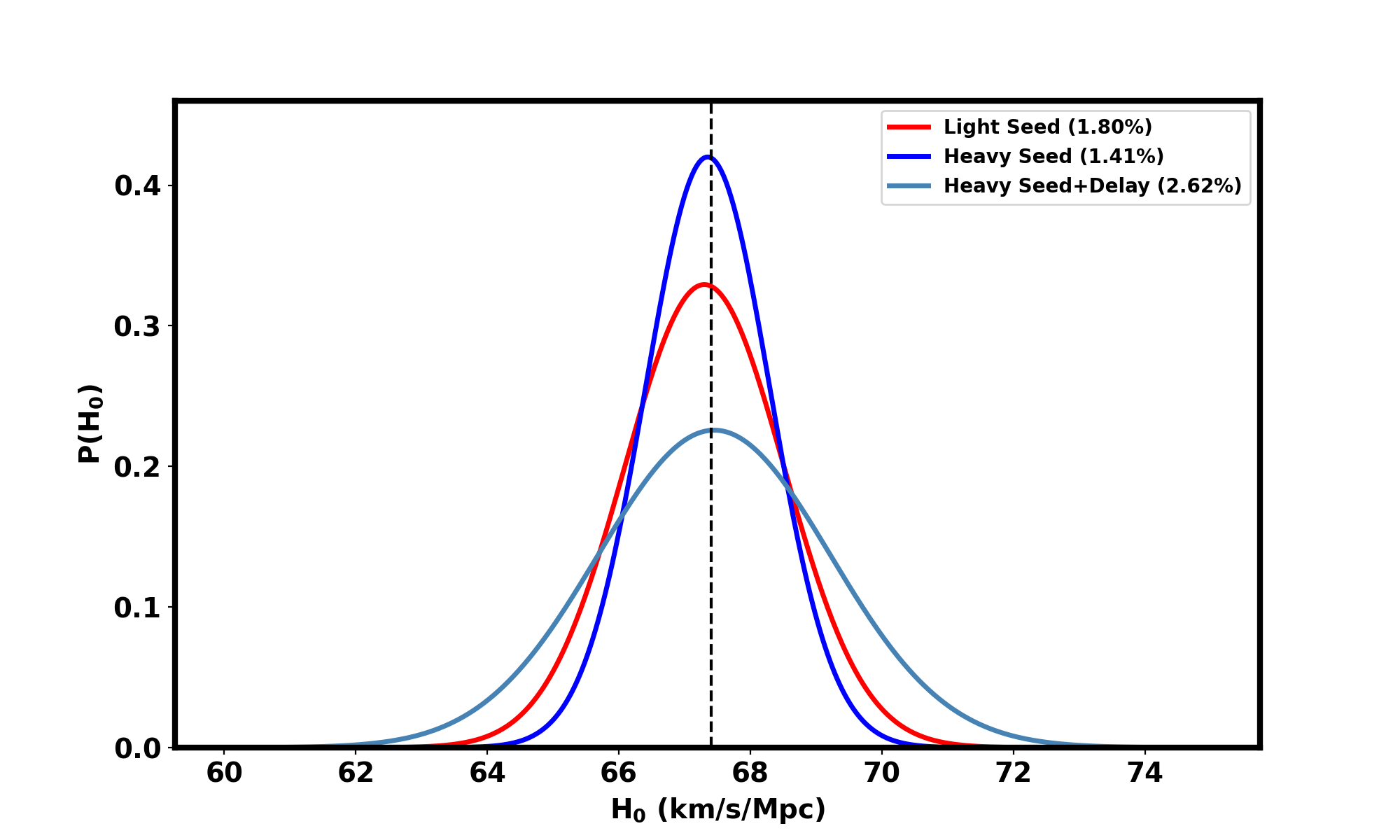}
    \caption{This plot shows the posterior distribution of the Hubble constant ($\mathrm{H_0}$) marginalized over $\mathcal{F}(z)$ for all events up to a redshift of 6, based on the $\mathrm{q^2}$ mass ratio distribution. The analysis uses the upcoming space-based LISA observatory, with observation periods of 4 years for the Light Seed and Heavy Seed models, and 10 years for the Heavy Seed+delay model. Results are shown assuming EM counterparts are detected for 75\% of the events. The 1$\sigma$ uncertainties are indicated in the legend.}
    \label{fig:H0posteriorq2}
\end{figure}

Figures \ref{fig:Fzreconstructionq2} and \ref{fig:H0posteriorq2} illustrate the results on the feasibility of inferring the parameter $\mathcal{F}(z)$ using the $\mathrm{q^2}$ mass ratio distribution with LISA. Figure \ref{fig:Fzreconstructionq2} presents the precision in reconstructing the redshift evolution of the Planck mass, $\mathcal{F}(z)$, highlighting LISA's measurement capabilities up to a redshift of 6, with reduced accuracy beyond a redshift of 3.0 due to limitations in BAO scale measurements. Figure \ref{fig:H0posteriorq2} shows the posterior distribution of the Hubble constant ($\mathrm{H_0}$) for events up to a redshift of $z=6.0$, assuming that EM counterparts are detected for 75\% of the events.

Here, we have used a power-law distribution proportional to $\mathrm{q}^{\alpha}$ with $\alpha = 2$. The measurements of $\mathrm{F(z)}$ and $\mathrm{H_0}$ will improve for any $\alpha > 2$, as the area under the curve of these distributions increases when $\mathrm{q}$ approaches 1. This means that the probability of obtaining nearly equal mass binaries is higher, leading to a higher SNR and, consequently, more events. Conversely, for $\alpha < 2$, the measurements become less accurate because the area under the curve decreases, reducing the likelihood of nearly equal mass binaries, resulting in a lower SNR and fewer events.

\clearpage

\bibliographystyle{JHEP.bst}
\bibliography{references}

\end{document}